\documentclass[aps,prb,showpacs,twocolumn]{revtex4-1}
\usepackage{bm,color,amsmath,amssymb,mathrsfs,latexsym,graphicx,psfrag}

\newcommand{\re}{\mathrm{Re}}

\def\ea{{\it et al.}}

\newcommand{\bk}{{\mathbf k}}

\newcommand{\br}{{\mathbf r}}

\newcommand{\be}{\begin{equation}}
\newcommand{\ee}{\end{equation}}

\def\be{\begin{equation}}
\def\ee{\end{equation}}
\def\bea{\begin{eqnarray}}
\def\eea{\end{eqnarray}}

\def\C60{A$_x$C$_{60}$}

\def\HgCu3{HgCa$_2$Cu$_3$O$_{8+y}$}
\def\HgCu4{HgBa$_2$Ca$_3$Cu$_4$O$_{10+y}$}
\def\TlCu{Tl$_2$Ba$_2$CuO$_{6+\delta}$}
\def\TlCu3{Tl$_2$Ba$_2$Ca$_2$Cu$_3$O$_{10+y}$}
\def\TlCu4{Tl$_2$Ba$_2$Ca$_3$Cu$_4$O$_{12+y}$}

\def\BiCu3{Bi$_2$Sr$_2$Ca$_{2}$Cu$_3$O$_y$}

\def\8LSCO{La$_{1.88}$Sr$_{.12}$CuO$_4$}
\def\110LNSCO{La$_{1.5}$Nd$_{0.4}$Sr$_{0.1}$CuO$_{4}$}
\def\stage4LCO{La$_{2}$CuO$_{4+\delta}$}
\def\Y248{YBa$_2$Cu$_4$O$_8$}

\def\NbSe2{NbSe$_2$}
\def\TaSe2{TaSe$_2$}
\def\TiSe2{TiSe$_2$}

\begin{document}

\title{Bulk Topological Invariants in Noninteracting Point Group Symmetric Insulators}
\author{Chen Fang$^{1}$, Matthew J. Gilbert$^{2,3}$,  B. Andrei Bernevig$^{1}$}
\affiliation{$^1$Department of Physics, Princeton University, Princeton NJ 08544}
\affiliation{$^2$Department of Electrical and Computer Engineering, University of Illinois,  Urbana IL 61801, USA}
\affiliation{$^3$Micro and Nanotechnology Laboratory, University of Illinois, 208 N. Wright St, Urbana IL 61801, USA}
\date{\today}

\begin{abstract}
We survey various quantized bulk physical observables in two- and three-dimensional topological band insulators invariant under translational symmetry and crystallographic point group symmetries (PGS). In two-dimensional insulators, we show that: (i) the Chern number of a $C_n$-invariant insulator can be determined, up to a multiple of $n$, by evaluating the eigenvalues of symmetry operators at high-symmetry points in the Brillouin zone; (ii) the Chern number of a $C_n$-invariant insulator is also determined, up to a multiple of $n$, by the $C_n$ eigenvalue of the Slater determinant of a noninteracting many-body system and (iii) the Chern number vanishes in insulators with dihedral point groups $D_n$, and the quantized electric polarization is a topological invariant for these insulators. In three-dimensional insulators, we show that: (i) only insulators with point groups $C_n$, $C_{nh}$ and $S_n$ PGS can have nonzero 3D quantum Hall coefficient and (ii) only insulators with improper rotation symmetries can have quantized magnetoelectric polarization $P_3$ in the term $P_3\mathbf{E}\cdot\mathbf{B}$, the axion term in the electrodynamics of the insulator (medium).
\end{abstract}
\maketitle

The study of novel topological phases of matter has become one of the most active fields in condensed matter physics. These phases are interesting because while deviating qualitatively from the conventional insulating phase, they cannot be described by any \emph{local} order parameter in the Ginzberg-Landau-Wilson spontaneous symmetry breaking paradigm. Heuristically, the word `topological' implies the presence of some \emph{global} property, i.e., contributed by all electrons in the system, that distinguishes this special phase. Such a property is usually marked by a global observable that takes different values in a topological phase and in the conventional insulating phase (normal phase) adiabatically continuable to the atomic limit. In addition, `topological' also implies that this global observable is quantized, or discretized, so that a topological phase cannot be adiabatically connected to the normal phase. Any quantized global observable characteristic to a topological phase is called a bulk topological invariant.

To date, topological phases have been identified in various systems, including those with intrinsic topological order (fractional quantum Hall states)\cite{laughlin1983,wen1995,Moore1991362}, topological band insulators\cite{qi2005,Kane:2005sf,Bernevig:2006kx,Fu:2006rm,Fu:2007fk,moore2007,koenig2007,zhang2009}, topological superconductors\cite{fu2008,Kitaev,schnyder2008} and topological semimetals\cite{SunK2011,Wan2011,XuG2011,Fang:2011,HalaszG2012}. The focus of this paper is on topological band insulators, or simply topological insulators (TI). A typical, and first, example of TI is the integer quantum Hall state\cite{Klitzing}. The bulk topological invariant of this state is the quantized Hall conductance, $\sigma_{xy}=ne^2/h$, where $n\in{Z}$, and its value remains fixed when magnetic field and gate voltage change within a certain range, resulting in a series of plateaus. Thouless \ea~\cite{Thouless:1982rz} showed that the quantized Hall conductance is proportional to the Chern number (TKNN number) of a $U(1)$ bundle over the 2D magnetic Brillouin zone (BZ), first time linking a quantized physical quantity to a topological number previously studied in the context of algebraic topology and differential geometry. It was then realized that since the Chern number is well-defined in any translationally invariant insulator regardless of the magnetic field, nonzero quantized Hall conductance can also appear \emph{without} external magnetic field (or net magnetic flux). Insulators with nonzero Chern numbers are later dubbed Chern insulators (or quantum anomalous Hall insulators)\cite{Haldane1988}. In general, the 2D Chern number of a translationally invariant insulator is given by\cite{Thouless:1982rz}
\bea\label{eq:defChern} C=\frac{i}{2\pi}\sum_{n\in{occ}}\int_{BZ}\epsilon_{ij}\partial_{i}\langle{u}_n(\bk)|\partial_{j}|u_n(\bk)\rangle{d}^2\bk,\eea where $|u_n(\bk)\rangle$ is the Bloch wavefunction of the $n$th band at $\bk$, and integral over BZ should be replaced with the one over magnetic BZ if there is an external magnetic field. From Eq.(\ref{eq:defChern}), it is clear that the Chern number is a global quantity contributed by all occupied electrons in the system.

If on top of translational symmetry, there is time-reversal symmetry (TRS) in the system, then one can prove that the Chern number defined in Eq.(\ref{eq:defChern}) always vanishes, a case often referred to as trivial. In the absence of nonzero Chern number, the natural question then is: what is the bulk topological invariant for this system, if any? The answer is obvious when the $z$-component of the total spin, $S^z_{tot}$ is conserved (commutes with the Hamiltonian), because in this case the two spin components are decoupled so that we can define the Chern numbers for spin up ($C_\uparrow$) and spin down ($C_\downarrow$) separately. Due to TRS, $C_\uparrow=-C_\downarrow$, hence the quantity\bea C_s=\frac{1}{2}(C_\uparrow-C_\downarrow)\eea is a nontrivial bulk topological invariant, also known as the spin Hall conductance\cite{Kane:2005sf,Bernevig:2006kx} (when multiplied by $\hbar$). However, when $S_{tot}^z$ is not conserved, the definition of the bulk topological invariant is not a simple extension of the Chern number. According to Fu and Kane\cite{Fu:2006rm,Fu:2007fk}, the nontrivial topology in a 2D insulator with translational symmetry and TRS can be characterized by a $Z_2$-number, i.e., a quantity that takes only two values. The explicit expression of this $Z_2$-number is not given here but we point out that it is linked to a quantized physical quantity, namely, the change in the `time-reversal polarization' over half of the BZ. The $Z_2$ bulk topological invariant thus defined can be generalized to 3D insulators with TRS\cite{Fu:2007fk,qi2008B,qi2011rev}, where it is given by
\bea\label{eq:defP3}P_3=\frac{1}{16\pi^2}\int{d^3k}\epsilon_{ijk}Tr[(\mathcal{F}_{ij}-\frac{2}{3}i\mathcal{A}_i\mathcal{A}_j)\mathcal{A}_k]\;\textrm{mod}\;1.\eea
In Eq.(\ref{eq:defP3}), we have defined the non-Abelian Berry connection \bea\mathcal{A}^{mn}_i(\bk)=i\langle{u}_m(\bk)|\partial_{i}|u_n(\bk)\rangle,\eea where $|u_n(\bk)\rangle$ is the Bloch state and the non-Abelian Berry curvature or field strength
\bea\mathcal{F}_{ij}(\bk)=\partial_{i}\mathcal{A}_j-\partial_{j}\mathcal{A}_i+i[\mathcal{A}_i,\mathcal{A}_j].\eea From Eq.(\ref{eq:defP3}), we can see that $P_3$ is a global quantity depending on the band structure of all occupied bands in the whole BZ. In 3D, $P_3$ has another physical meaning: it is the coefficient of the axion term of the electrodynamics of the insulator in the compact space (having periodic boundary conditions in all directions)\cite{qi2011rev,Qi:2008sf},
\bea H_{axion}=P_3\mathbf{E}\cdot\mathbf{B},\eea and is in principle measurable with electromagnetic induction experiments. In order for $P_3$ to be a $Z_2$-index (taking only two possible values), it must be quantized by some symmetry. In an insulator with no symmetry, $P_3$ defined in Eq.(\ref{eq:defP3}) is an arbitrary number between zero and unity; but TRS quantizes it to either zero or one-half. Nontrivial $Z_2$ topological insulators in 2D and 3D have odd number of gapless Dirac edge and surface modes respectively, as has been confirmed by experiments\cite{Hsieh:2008fk,hsieh2009a,Hsieh2009}. Other than the linear dispersion around the Dirac points, these surface/edge states distinguish themselves by the absence of backscattering\cite{koenig2007,liu2009,ZhouX2009}.

From these previous studies, we can see that symmetries of a system play a key role in the definition of its bulk topological invariants. TRS has a two-fold effect: it on one hand constrains the Chern number to be zero, and on the other hand makes possible the definition of a new $Z_2$ number. An interesting question then is: if we consider adding other symmetries to an insulator, will there be new bulk topological invariant(s) characteristic to the added symmetries? Answering this question on a general ground is not easy, because it takes a detailed study of the homotopy groups of continuous mappings from the $d$-dimensional BZ to a target space of projectors under the symmetry constraints. Along this line, Schnyder \ea~ gave a list of all bulk topological invariants for each of the ten classes of Hamiltonians in the presence/absence of TRS, particle-hole symmetry and chiral symmetry for $d=1,2,3$\cite{schnyder2008}.

All the three symmetries considered in Ref.\onlinecite{schnyder2008} are symmetries on internal degrees of freedom, which do not change the position or orbital character of an electron. Topological properties protected by these internal symmetries are in general robust against random factors such as disorder. Nevertheless, the requirement of TRS excludes a spectrum of interesting materials that have magnetism while particle-hole symmetry and chiral symmetry are not to be found in realistic normal insulators but only in superconductors or maybe optical lattices. Therefore, it is desired to extend the study of topological invariants to systems with other symmetries. In condensed matter electronic systems, it is natural to consider the lattice symmetries in the classification of insulators, because of their universal existence in solids, nonmagnetic and magnetic. Along this line, Fu first studied possible new $Z_2$-index in 3D spinless TRS insulators with fourfold rotation symmetry\cite{Fu:2011}; Hughes \emph{et al}\cite{hughes2010inv} and Turner\emph{et al}\cite{Turner:2012} independently developed theories for inversion symmetric topological insulators without TRS.

In 3D there are 230 types of lattice symmetries, in which the simplest, yet nontrivial, one is the inversion symmetry, which sends an electron at $\br$ to $-\br$ leaving its spin invariant. Insulators with inversion symmetry have been previously studied\cite{fu2007a,hughes2010inv,Turner:2012} and it is shown that this symmetry affects the topological properties in two ways. First, its existence places constraints on the values of bulk topological invariants such as the Chern number and the $Z_2$ number: in Ref.\onlinecite{hughes2010inv} it is shown that the parity (even or odd) of the Chern number is determined by the inversion eigenvalues of occupied bands at four inversion invariant $\bk$-points in the BZ, regardless of the Bloch wavefunction at all other $\bk$-points; the $Z_2$ number of a TI with TRS can also be determined by the inversion eigenvalues at all inversion invariant $\bk$-points\cite{fu2007a}. Second, inversion symmetry alone brings about new topological invariants \emph{without} TRS. We have mentioned that $P_3$ is quantized to zero and one-half by TRS, and Ref.\onlinecite{hughes2010inv,Turner:2012} shows that in the absence of TRS, $P_3$ is still quantized by inversion symmetry and is equal to, up to an integer, half of the winding number of the inversion sewing matrix (to be defined in Sec.\ref{sec:prelim}(B)).

In this paper, we study topological insulators subject to a more general class of lattice symmetries: crystallographic point group symmetries, or simply point group symmetries (PGS)\cite{Fu:2011,Fu:2012}. Insulators with PGS are invariant under a certain set of rotations and reflections that leave at least one point fixed in space. In the language of crystallography, by only considering PGS and lattice translational symmetry, we restrict the discussion to insulators with symmorphic space groups. Simply speaking, in a symmorphic space group, there are only point group operations plus translations by lattice vectors, while the a general space group can contain combined operations of a point group operation followed by a translation of fractions of lattice vectors. We ask two questions regarding the relation between PGS and bulk topological invariants: (i) how does a given PGS constrain the value of a topological invariant such as the Chern number? (ii) can a given PGS give us new bulk topological invariants, or quantized global quantities?

For 2D insulators, we answer these questions by an exhaustive discussion of all nine nontrivial PGS invariant insulators, divided into two types: those invariant under cyclic PGS $C_{n=2,3,4,6}$ and those under dihedral PGS $D_{n=1,2,3,4,6}$. For $C_n$-invariant insulators, we first show that the Chern number is determined by, or constrained by, the $C_m$ eigenvalues of each occupied single particle band at discrete high-symmetry points in BZ, where $m$ evenly divides $n$. From there we further prove that the Chern number modulo $n$ is determined by the $C_n$ eigenvalue of the \emph{many-body} ground state, or Slater determinant in the noninteracting case. For $D_n$-invariant insulators, we prove that while the Chern number is constrained to zero, the electric polarization, or more strictly, the dislocation of the electronic charge center from the nearest lattice point, becomes a bulk topological invariant, the value of which is determined by eigenvalues of $C_m$.

In 3D, there are in total 31 nontrivial PGS, and instead of exercising an exhaustive study, we focus on establishing a link between the given PGS and the presence/absence of two experimentally interesting topological invariants: nonzero quantized 3D Hall conductance and quantized magnetoelectric coefficient $P_3$. We find that only $C_n$, $C_{nh}$, $S_{n}$ PGS are compatible with nonzero Hall conductance while all other PGS lead to zero Hall conductance in all three components. Moreover, we prove that the sufficient and necessary condition for the quantization of $P_3$ (to zero and one-half) is that the PGS contains at least one improper rotation, and $2P_3$ can be expressed as the winding number of the sewing matrix associated with that improper rotation symmetry. This excludes the following point groups from having quantized $P_3$: $C_n$, $D_n$, $O$, $T$, 11 distinct point groups in total. These major results are summarized in Table \ref{tab:2D_table} for 2D and Table \ref{tab:3D_table} for 3D.

\begin{table}[!htb]
\includegraphics[width=8cm]{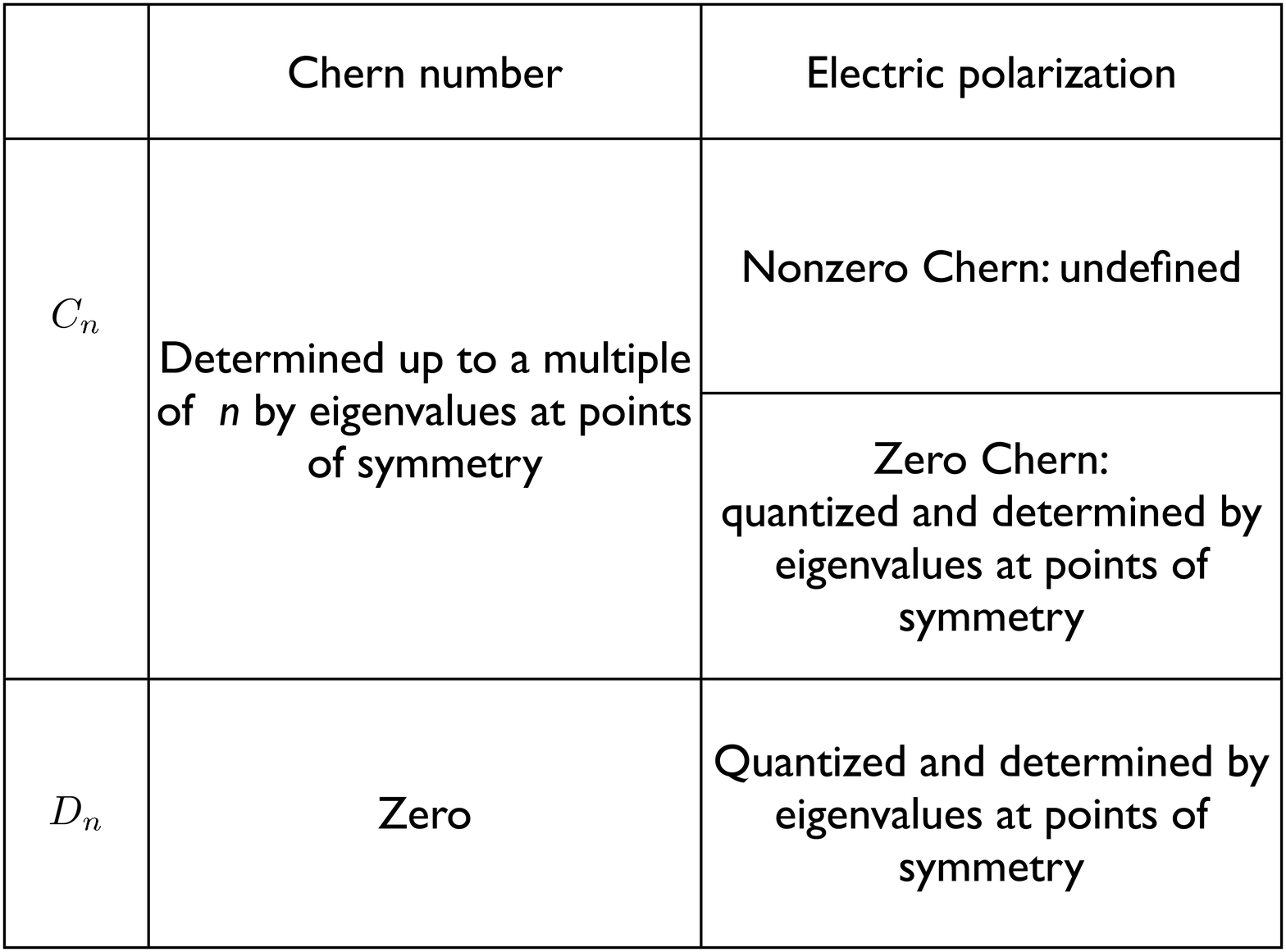}
\caption{Summary of major results regarding 2D PGS insulators. The index $n$ always takes values of $n=2,3,4,6$; and `eigenvalues at points of symmetry' is the short for $C_m$-eigenvalues at $C_m$ invariant points, where $m$ divides $n$.}\label{tab:2D_table}
\end{table}

\begin{table}[!htb]
\includegraphics[width=8cm]{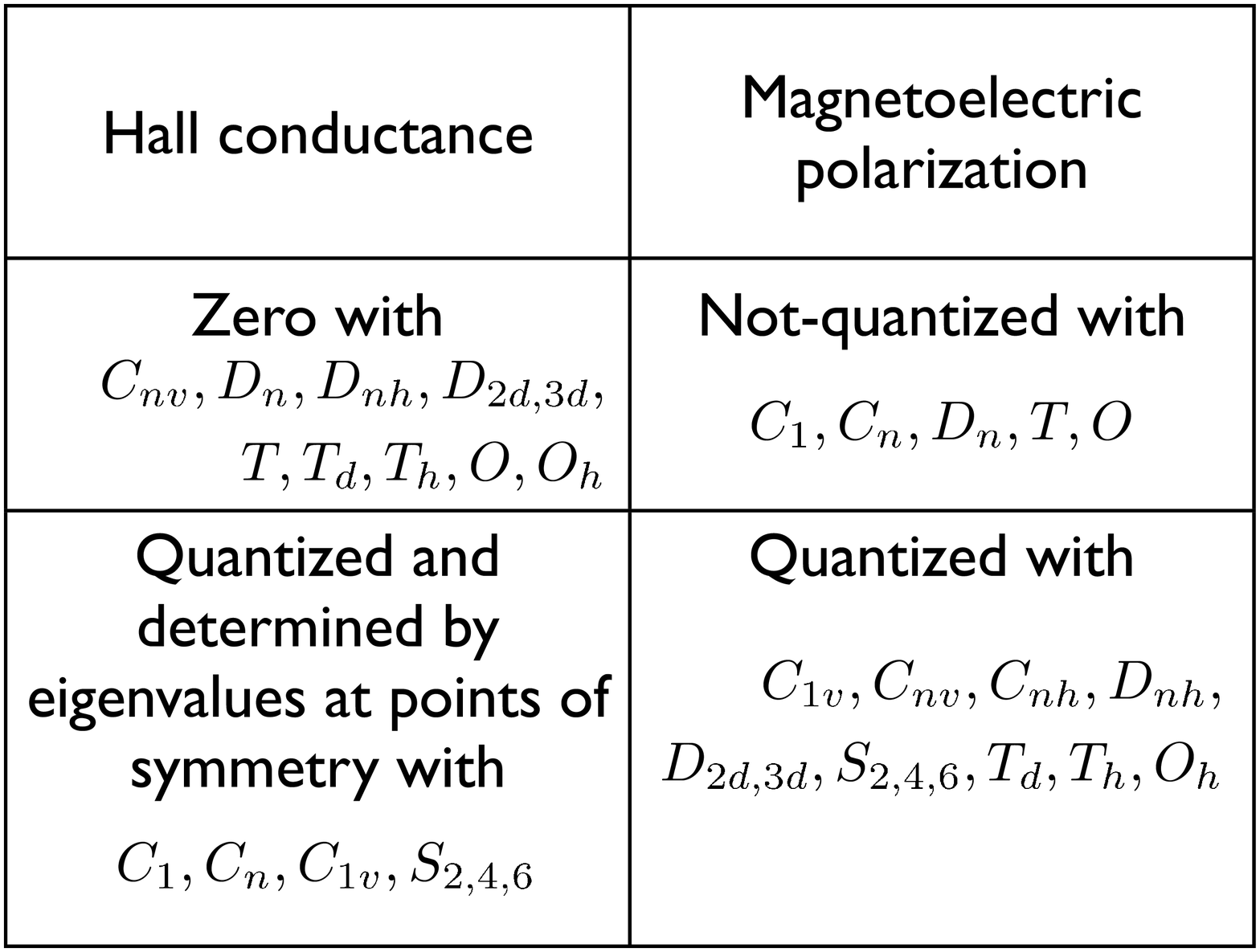}
\caption{Summary of major results regarding 3D PGS insulators.}\label{tab:3D_table}
\end{table}

The paper is organized as follows. In Sec.\ref{sec:prelim}, we introduce concepts and basic formulae constantly used throughout the paper, including a brief introduction to Hilbert space representations of crystallographic point groups in Sec.\ref{sec:prelim}(A), definition and basic properties of the sewing matrix associated with a PGS operator in Sec.\ref{sec:prelim}(B) and the definition of basic properties of path-ordered loop integrals in BZ in Sec.\ref{sec:prelim}(C). In Sec.\ref{sec:pgs2d}, we show how the Chern numbers in $C_n$-invariant 2D insulators can be determined by eigenvalues of symmetry operators at high-symmetry $\bk$-points in Sec.\ref{sec:pgs2d}(A), and how the Chern number in $C_n$-invariant 2D insulators can be determined by the $C_n$ eigenvalue of the Slater determinant in Sec.\ref{sec:pgs2d}(B). In Sec.\ref{sec:diheral}, we show that $D_n$-invariant insulators have zero Chern number in Sec.\ref{sec:diheral}(A) and that the electric polarizations are quantized thus being topological invariants and how to calculate them in Sec.\ref{sec:diheral}(B). We study 3D insulators with PGS in Sec.\ref{sec:pgs3d}, including the 3D quantum Hall effects in Sec.\ref{sec:pgs3d}(A) and quantization of 3D magnetoelectric coefficient $P_3$ in Sec.\ref{sec:pgs3d}(B). In Sec.\ref{sec:conclusion} we conclude the work.

\section{Preliminaries}
\label{sec:prelim}
Before proceeding with the relevant background material, we briefly clarify the notations we use throughout the paper. When used as indices, the Greek letters will always span the orbitals, while Roman letters denote either the bands or spatial directions ($x,y,z$ or $k_x,k_y,k_z$).

Throughout the text there are three different linear spaces in which operators are considered. First is the single-particle Hilbert space. The dimension of the single particle Hilbert space is $N\times{N}_{orb}$, where $N$ is the total number of sites and $N_{orb}$ total number of atomic orbitals, including spin, per site. The second linear space is the orbital space. The dimension of the orbital space is $N_{orb}$. The third linear space is the occupied-band space. The dimension of this subspace is $N_{occ}$, where $N_{occ}<N_{orb}$ is the number of occupied bands.

Summation of repeated indices is implicitly indicated when and only when the summation is over an index denoting the Cartesian coordinates: $x,y,z$ and $k_x,k_y,k_z$.

\subsection{Single-valued and double-valued representations of a double point group}

In physics, properties of Hilbert space representations, i.e., operators, of symmetry group are more relevant than those of the group itself. For a point group, before defining the operators corresponding to each symmetry operation,
it is necessary to extend the point group, $G$, to the double-point group $G^D$ defined as $G^D=G\cup(\bar{E}*G)$, where $\bar{E}$ is a rotation of angle $2\pi$ about any axis in space, which satisfies $\bar{E}^2=E$, and $\bar{E}g=g\bar{E}$ for $g\in{G}$. In this definition, we notice that a rotation of angle $2\pi$ is distinct from the identity element in $G^D$. All representations of $G^D$ can be classified into single-valued representations and double-valued representations, in which $D(E)=D(\bar{E})$ and $D(E)=-D(\bar{E})$ respectively. In a physical system, the one-particle Hilbert space representation of $G^D$ is a single-valued representation when the particle is spinless or of integer spin, and is a double-valued representation when the particle is of half odd integer spin. For example of the $C_n$ point group, if $\hat{C}_n$ is the single particle operator for the $n$-fold rotation (and we will use a hat to denote operators in the single particle Hilbert space), then we have $\hat{C}_n^n=\hat{I}$ if the particle is spinless or of integer spin, and $\hat{C}_n^n=-\hat{I}$ if the particle is of half odd integer spin, where $\hat{I}$ is the identity operator. To put the relations in a single equation: $\hat{C}_n^n=(-1)^F\hat{I}$ where $F$ is two times the total spin of the particle.

\subsection{Sewing matrix associated with a point group symmetry}

It is assumed that our system is defined on a lattice, has translational invariance and is noninteracting:
\bea\hat{H}&=&\sum_{\bk\in{BZ}}\hat{h}(\bk)\\
\nonumber&\equiv&\sum_{\alpha,\beta,\bk\in{BZ}}\tilde{H}^{\alpha\beta}(\bk)c^\dag_\alpha(\bk)c_\beta(\bk),\eea where $\alpha,\beta$ denote the orbitals within a unit cell. When we say a Hamiltonian is invariant under point group $G$, we mean given $R\in{G}$, there is
\bea\label{eq:Rinvariance}\hat{R}\hat{h}(\bk)=\hat{h}(R\bk)\hat{R},\eea where $R\bk$ is the transformed wavevector $\bk$ under $R$. For example, if $R$ is an $n$-fold rotation $C_n$ about $z$-axis, $C_n\bk=(k_x\cos(2\pi/n)-k_y\sin(2\pi/n),k_x\sin(2\pi/n)+k_y\cos(2\pi/n),k_z)$; if $R$ is a mirror reflection about the $xy$-plane $M_{xy}$, $M_{xy}\bk=(k_x,k_y,-k_z)$.

At each $\bk$, the eigenstates of $\hat{h}(\bk)$ are called the Bloch states, whose annihilation operators $\gamma_n(\bk)$'s satisfy
\bea\label{eq:defsingle}[\gamma_n(\bk),\hat{h}(\bk)]=E_n(\bk)\gamma_n(\bk),\eea where $n=1,...,N_{orb}$ is the band index. Then one can arrange $\gamma_n$ in the ascending order of $E_n(\bk)$, such that for a fully gapped insulator, we have $E_n(\bk)<E_f$ for $n<N_{occ}\le N_{orb}$ for all $\bk\in{BZ}$. Combining Eq.(\ref{eq:Rinvariance}) and Eq.(\ref{eq:defsingle}), we have
\bea[\hat{R}\gamma_n(\bk)\hat{R}^{-1},\hat{h}(R\bk)]=E_n(\bk).\eea This equation shows that $\hat{R}\gamma_n(\bk)\hat{R}^{-1}$ is a Bloch state operator at $R\bk$ with the same energy. Therefore in general,
\bea\label{defSewing}\hat{R}\gamma_n(\bk)\hat{R}^{-1}=\sum_{m\in{occ}}(\mathcal{B}_R(\bk))_{mn}\gamma_m(R\bk),\eea where $m\in{occ}$. Considering degeneracies, $(\mathcal{B}_R(\bk))_{mn}$ is \emph{not} proportional to $\delta_{mn}$. The $N_{occ}$-by-$N_{occ}$ matrix $\mathcal{B}_R(\bk)$ is called the sewing matrix associated with PGS operator $R$, and we will keep using curly letters to represent matrices in the space of occupied bands. Defining $|\psi_n(\bk)\rangle=\gamma^\dag_n(\bk)|0\rangle$, where $|0\rangle$ is the vacuum state with no fermion, we can also put the sewing matrix definition in the form
\bea\label{eq:defSewing2}(\mathcal{B}_R(\bk))_{mn}=\langle{\psi}_m(R\bk)|\hat{R}|\psi_n(\bk)\rangle,\eea for $m,n\in{occ}$. (Throughout the paper, when it is a Greek letter inside $|\rangle$/$\langle|$, the symbol denotes a column/row vector in the single particle Hilbert space.)

An alternative and useful expression of the sewing matrix can be found by noticing that each Bloch state $|\psi_n(\bk)\rangle$ is the direct product of a plane wave $e^{i\bk\cdot\br}$ and an orbital part $|u_n(\bk)\rangle$, and noticing that the rotation operator $\hat{R}$ acts on the two parts separately sending $e^{i\bk\cdot\br}$ to $e^{iR\bk\cdot\br}$ and $|u_n(\bk)\rangle$ to $\tilde{R}|u_n\rangle$:
\bea(\mathcal{B}_{R}(\bk))_{mn}=\langle u_m(R\bk)|\tilde{R}|u_n(\bk)\rangle,\eea , where $\tilde{R}$ and all capital letters with a tilde are matrices in the orbital space and when a Roman letter appears in $|\rangle$/$\langle|$, the symbol denotes a column/row vector in the orbital space.

For a given PGS operator $R$, there are points in BZ at which $\bk_{inv}=R\bk_{inv}$. These points are called high-symmetry points, or points of symmetry. At $\bk_{inv}$, from Eq.(\ref{eq:Rinvariance}), we have
\bea[\hat{R},\hat{h}(\bk_{inv})]=0.\eea Therefore, we can find a common set of eigenstates of both $\hat{R}$ and $\hat{h}(\bk_{inv})$. In that basis, the sewing matrix is diagonal
\bea\label{eq:diagonalB}(\mathcal{B}_R(\bk_{inv}))_{mn}=R_m(\bk)\delta_{mn},\eea where $R_m$ is the eigenvalue of $\hat{R}$ on the $m$th band. The determinant of the sewing matrix at $\bk_{inv}$ can then be calculated as, using Eq.(\ref{eq:diagonalB}),
\bea\label{eq:detB}\det[\mathcal{B}_R(\bk_{inv})]=\prod_{n\in{occ}}R_n(\bk_{inv}).\eea Since the determinant is independent of the choice of basis, $\det[\mathcal{B}_R(\bk_{inv})]$ is a gauge invariant quantity.

\subsection{Monodromy}

Path-ordered loop integrals and line integrals of exponentiated Berry connection are extensively used in Sec.\ref{sec:pgs2d}(A). Here we briefly develop basic properties of these integrals. Alternative formulas can be found in Ref.[\onlinecite{Alexandradinata:2012}].

A path-ordered loop integral of Berry connection describes a unitary evolution of the linear subspace spanned by the $N_{occ}$ eigenstates of $\tilde{H}(\bk)$ around a closed circuit in the BZ. Mathematically, it is expressed in terms of the non-Abelian Berry connection\bea\mathcal{W}_L=P\exp(i\oint_L\mathcal{A}(\bk)\cdot d\bk),\eea where $L$ means a closed loop and $P$ means `path ordered'. Here we point out that in 2D, the determinant of this integral
\bea\det(\mathcal{W}_L)&=&\det(P\exp(i\oint_L\mathcal{A}(\bk)\cdot d\bk))\\
\nonumber&=&\det(\exp(i\oint_L\mathcal{A}(\bk)\cdot d\bk))\\
\nonumber&=&\exp(iTr(\oint_L\mathcal{A}(\bk)\cdot d\bk))\\
\nonumber&=&\exp(i\phi_{B}),\eea where $\phi_{B}$ is the Berry phase associated with the loop.

In Section \ref{sec:pgs2d}(A), we will calculate the determinants of loop integrals that encircle a portion of BZ. To calculate them, it is convenient to define the Wilson line between the occupied-subspaces at $\bk_1$ and $\bk_2$, obtained by exponentiating the Berry connection:
\bea\mathcal{U}_{\bk_1\bk_2}=P\exp(i\int_{\bk_2}^{\bk_1}\mathcal{A}(\bk)d\bk).\eea In a finite system with discrete $\bk$, we have, equivalently,
\bea
&&(\mathcal{U}_{\bk_1\bk_2})_{mn}=\\
\nonumber&&\sum_{a,b...}\langle{u}_{m}(\bk_1)|u_a(\bk'_1)\rangle\langle{u}_a(\bk'_1)|{u}_b(\bk'_2)\rangle\langle{u}_b(\bk'_2)|...|u_n(\bk_2)\rangle,\eea where $\bk'_1,\bk'_2...$ form a path connecting $\bk_1$ and $\bk_2$.

This connection is by definition an $N_{occ}\times N_{occ}$ matrix. We can also define an orbital space operator $\tilde{U}_{\bk_1\bk_2}$ associated with this matrix:
\bea\tilde{U}_{\bk_1\bk_2}&=&\sum_{i,j\in{occ}}(\mathcal{U}_{\bk_1\bk_2})_{ij}|u_i(\bk_1)\rangle\langle{u}_j(\bk_2)|\\
\nonumber&=&P\exp(i\int_{\bk_2}^{\bk_1}\tilde{P}_\bk\mathbf{\partial}\tilde{P}_\bk\cdot{d}\bk),\eea where $\tilde{P}_\bk=\sum_{i\in{occ}}|u_i(\bk)\rangle\langle{u}_i(\bk)|$ is the projector onto the occupied subspace at $\bk$. In a finite size system, alternatively, one has
\bea\tilde{U}_{\bk_1\bk_2}=\tilde{P}_{\bk_1}(\prod_{j=1,2,...}\tilde{P}_{\bk'_j})\tilde{P}_{\bk_2}.\eea The advantage of using $\tilde{U}_{\bk_1\bk_2}$ is its gauge invariance as $\tilde{P}_\bk$ only depends on the Hamiltonian. If the Hamiltonian has a symmetry $R$, it can be proven that \bea\label{eq:transformU}\tilde{R}\tilde{U}_{\bk_1\bk_2}\tilde{R}^{-1}=\tilde{U}_{R\bk_1R\bk_2},\eea although no simple relation exists for their corresponding matrices in the occupied band space $\mathcal{U}_{\bk_1\bk_2}$ and $ \mathcal{U}_{R\bk_1R\bk_2}$. This property will be extensively used in the monodromy proof provided in Section \ref{sec:pgs2d}(A).

\section{Two-dimensional Insulators with Cyclic Point Group Symmetries}
\label{sec:pgs2d}

There are in total 10 point groups in 2D. One is the trivial group containing only the identity, and among the nontrivial 9 PGS, there are four cyclic PGS denoted by $C_{n=2,3,4,6}$. The symmetry group $C_n$ is generated by an $n$-fold rotation about the $z$-axis that is out-of-plane. In this section we will study $C_n$-invariant insulators in two dimensions.

\subsection{Chern number in terms of rotation eigenvalues at high-symmetry points}

In the following, we endeavor to prove in a $C_n$-invariant insulator, the Chern number modulo $n$ is given by $C_m$ eigenvalues of all occupied bands at each $C_m$-invariant $\bk$-point for each $m$ dividing $n$. The case of $n=2$ has been studied in Ref.\onlinecite{hughes2010inv} and the result is \bea\label{eq:ChernC2}(-1)^{C}=\prod_{i\in{occ.}}\zeta_i(\Gamma)\zeta_i(X)\zeta_i(Y)\zeta_i(M),\eea where $\zeta_i(\bk=\Gamma,X,Y,M)$ is the eigenvalue of the operator $\hat{C}_2$ at $C_2$-invariant $\mathbf{k}$-point on the $i$th band. This equation shows that the parity (even or odd) of the Chern number can be determined by knowing the inversion eigenvalues at four high-symmetry points.

Now we move beyond the simple and already discussed case of $C_2$ symmetric systems to calculate the Chern number in  $C_{n}$-invariant insulators for $n=3,4,6$. We will calculate the Chern number by relating it to the determinant of a special closed loop integral of exponentiated Berry connection defined in BZ. In the main text the detailed calculation is only shown for $C_4$-invariant insulators; for $C_{3,6}$-invariant insulators, the main conclusions are stated here while details are provided in Appendix \ref{apndx:mono}.
\begin{figure}[t]
\centering
\includegraphics[width=8cm]{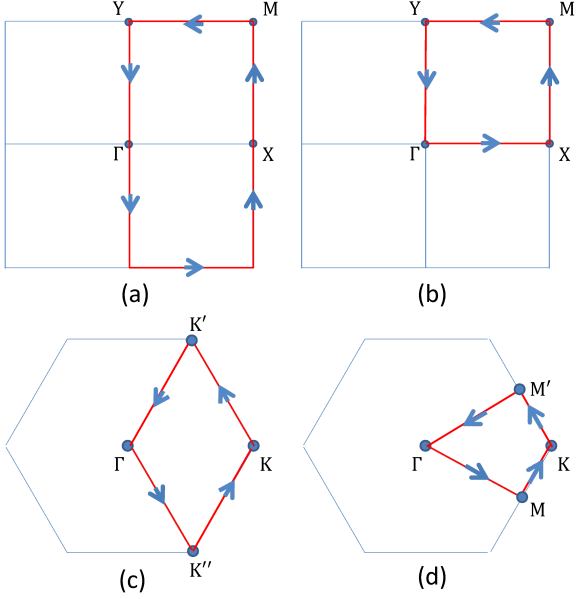}
\caption{Schematic of the loop integrals used in the monodromy proof of the relationship between the Chern number and the eigenvalues at high symmetry points within the Brillouin zone in $C_n$ invariant topological insulators with $n=2,4,3,6$ in (a), (b),(c), and (d) respectively.}\label{fig:loops}
\end{figure}

For $n=4$, the loop is chosen as $\lambda=\Gamma{X}{M}{Y}\Gamma$ shown in Fig.\ref{fig:loops}(b), and the determinant of the path-ordered loop integral becomes \bea\label{eq:OneFourthLoop}\det[P\exp(\oint_{\lambda}{\mathbf{\mathcal{A}(k)}}\cdot d\mathbf{k})]=e^{i\pi{C}/2}.\eea To prove this we note that due to $C_4$, we have the following symmetry in the field strength \bea Tr[\mathcal{F}_{xy}(k_x,k_y)]=Tr[\mathcal{F}_{xy}(-k_y,k_x)]\eea (for a proof see Appendix \ref{apndx:transformF}) and hence the integral of Berry phase inside $\lambda$ is one fourth of the Berry phase in the whole BZ ($2\pi{C}$). On the other hand, the loop integral is given by the connection matrices defined in Sec.\ref{sec:prelim}(C) \bea\mathcal{W}_{\lambda}=\mathcal{U}_{\Gamma{X}}\mathcal{U}_{XM}\mathcal{U}_{MY}\mathcal{U}_{Y\Gamma}.\eea With $C_4$ symmetry at hand and utilizing the periodicity of the BZ, we notice that $\tilde{C}_4\tilde{U}_{\Gamma{Y}}\tilde{C}_4^{-1}=\tilde{U}_{\Gamma{X}}$ and $\tilde{C}^{-1}_4\tilde{U}_{YM}\tilde{C}_4=\tilde{U}_{MX}$. We may further simplify the determinant as (see Appendix \ref{apndx:Eq7} for more additional details)\begin{widetext}\bea \det(\mathcal{W}_{\lambda})&=&\det(\mathcal{B}_{C_4}(\Gamma)\mathcal{U}_{\Gamma{Y}}\mathcal{B}_{C_2}(Y)^{-1}\mathcal{U}_{YM}\mathcal{B}_{C_4}(M)\mathcal{U}_{MY}\mathcal{U}_{Y\Gamma})\\
\nonumber&=&\det(\mathcal{B}_{C_4}(\Gamma)\mathcal{B}_{C_2}^{-1}(Y)\mathcal{B}_{C_4}(M))\det(\mathcal{U}_{\Gamma Y}\mathcal{U}_{YM}\mathcal{U}_{MY}\mathcal{U}_{Y\Gamma})\\
\nonumber&=&\det(\mathcal{B}_{C_4}(\Gamma)\mathcal{B}_{C_2}^{-1}(Y)\mathcal{B}_{C_4}(M)),\label{eq:NumberFromThreePoints}\eea\end{widetext} where we have used $\mathcal{U}_{YM}=\mathcal{U}^{-1}_{MY}$. From Eq.(\ref{eq:OneFourthLoop},\ref{eq:NumberFromThreePoints}) and Eq.(\ref{eq:detB}), we obtain\bea i^{C}=\prod_{n\in{occ.}}\xi_n(\Gamma)\xi_n(M)\zeta_n^{-1}(Y)\label{eq:temp1},\eea where $\xi_i(\bk)$ is the eigenvalue of $\hat{C}_4$ at the $C_4$-invariant $\bk$-point on the $i$th band. Since $\hat{C}_2^2=(-1)^F$, we have $\zeta^2(\bk=X,Y)=(-1)^F$ or
\bea\label{eq:zeta}\zeta^{-1}(Y)=(-1)^F\zeta(Y).\eea Substituting Eq.(\ref{eq:zeta}) into Eq.(\ref{eq:temp1}), we obtain
\bea i^{C}=\prod_{i\in{occ.}}(-1)^{F}\xi_i(\Gamma)\xi_i(M)\zeta_i(Y)\label{eq:ChernC4}.\eea Eq.(\ref{eq:ChernC4}) tells us that given the eigenvalues of $\hat{C}_4$ at $\Gamma,M$ and those of $\hat{C}_2$ at $Y$, one can determine the Chern number up to a multiple of four. It may be disconcerting that in Eq.(\ref{eq:ChernC4}) we only include the $C_2$ eigenvalues at $Y$ but not at $X$, but in fact they are identical: in systems with $C_4$ symmetry, $\zeta_i(0,\pi)=\zeta_i(\pi,0)$ as the two points are related by $C_4$ rotation.

For $n=3$ and $n=6$, the proof takes very similar steps which we leave to Appendix \ref{apndx:mono}, and only differs in that one chooses the loops as shown in Fig.\ref{fig:loops}(c) and Fig.\ref{fig:loops}(d), respectively. We simply quote the salient results here\bea e^{i2\pi{C}/3}&=&\prod_{i\in{occ.}}(-1)^{F}\theta_i(\Gamma)\theta_i(K)\theta_i(K')\label{eq:ChernC3},\\
e^{i\pi{C}/3}&=&\prod_{i\in{occ.}}(-1)^{F}\eta_i(\Gamma)\theta_i(K)\zeta_i(M)\label{eq:ChernC6},\eea where $\theta_i(\bk)$ and $\eta_i(\bk)$ represent the eigenvalues of $\hat{C}_3$ and $\hat{C}_6$ at $C_3$- and $C_6$-invariant $\bk$-points, respectively.

\subsection{Chern number in terms of the $C_n$ eigenvalue of the Slater determinant}

We have shown how the Chern number is related to symmetry eigenvalues of the single particle states at high-symmetry points. Now we will see that another connection between the Chern number and $C_n$-invariance presents itself as one considers the $C_n$ eigenvalue of the \emph{many-body} ground state (under the assumption that this is non-degenerate), or the Slater determinant state of a $C_n$-invariant insulator:
\bea\label{eq:ChernSlater}\exp(iC\frac{2\pi}{n})=f_n(N)\rho_n,\eea where $f_n(N)$ is either $+1$ or $-1$ depending on the total number of sites and $\rho_n$ is the eigenvalue of $\hat{C}_n$ of the Slater determinant state $|\Phi_0\rangle$, i.e., $\hat{C}_n|\Phi_0\rangle=\rho_n|\Phi_0\rangle$. Eq.(\ref{eq:ChernSlater}) can be proved by expressing $\rho_n$ in terms of the eigenvalues of the single particle states at high-symmetry points and then using Eq.(\ref{eq:ChernC2},\ref{eq:ChernC3},\ref{eq:ChernC4},\ref{eq:ChernC6}) to relate $\rho_n$ to the Chern number for $n=2,3,4,6$.

We begin by looking at the case of $C_2$-invariant insulators. The Slater determinant of the ground state is given by\bea|\Phi_0\rangle=\prod_{\bk\in{BZ}}\prod_{i\in{occ.}}\gamma^\dag_i(\mathbf{k})|0\rangle.\eea We can pair $\mathbf{k}$ and $-\mathbf{k}$ and this results in only four $\mathbf{k}$'s that cannot be paired as they are inversion invariant. So the Slater determinant can be written as\bea|\Phi_0\rangle&=&\prod_{i\in{occ.}}\gamma^\dag_i(\Gamma)\gamma^\dag_i(Y)\gamma^\dag_i(X)\gamma^\dag_i(M)\\
\nonumber&&\times\prod_{\mathbf{k}\in{BZ/2}}(\gamma^\dag_i(\mathbf{k})\gamma^\dag_i(-\mathbf{k}))|0\rangle.\eea Note that in writing down the above equation, and throughout the rest of the Section, we have implicitly assumed that all high-symmetry points exist in the finite system, and for a discussion of all other cases, see Appendix \ref{apndx:points}. Now consider the action of $\hat{C}_2$ on this state:\begin{widetext}\bea\label{eq:temp2}\hat{C}_2|\Phi_0\rangle&=&\prod_{i\in{occ.}}(\hat{C}_2\gamma^\dag_i(\Gamma)\hat{C}_2^{-1})(\hat{C}_2\gamma^\dag_i(Y)\hat{C}_2^{-1})(\hat{C}_2\gamma^\dag_i(X)\hat{C}_2^{-1})(\hat{C}_2\gamma^\dag_i(M)\hat{C}_2^{-1})\\\nonumber&&\prod_{\mathbf{k}\in{BZ/2}}(\hat{C}_2\gamma^\dag_i(\mathbf{k})\hat{C}_2^{-1}\hat{C}_2\gamma^\dag_i(-\mathbf{k})\hat{C}_2^{-1})|\Phi_0\rangle\\
\nonumber&=&(-1)^{N_{occ}(N-4)/2}\det(\mathcal{B}_{C_2}(\Gamma))\det(\mathcal{B}_{C_2}(X))\det(\mathcal{B}_{C_2}(Y))\det(\mathcal{B}_{C_2}(M)) \prod_{\mathbf{k}\in{BZ/2}}\det(\mathcal{B}_{C_2}(\mathbf{k})\mathcal{B}_{C_2}(-\mathbf{k}))|\Phi_0\rangle.\eea \end{widetext} Using the sewing matrix property to see that $\mathcal{B}_{C_2}(\mathbf{k})=(-1)^{F}\mathcal{B}_{C_2}^{-1}(-\mathbf{k})$, one obtains from Eq.(\ref{eq:temp2}) \bea\label{eq:Slater_C2}\nonumber\hat{C}_2|\Phi_0\rangle=(-1)^{\frac{(F-1)N_{occ}(N-4)}{2}}\prod_{i\in{occ.}}\zeta_i(\Gamma)\zeta_i(X)\zeta_i(Y)\zeta_i(M)|\Phi_0\rangle.\\\eea
Combining Eq(\ref{eq:Slater_C2}) and Eq(\ref{eq:ChernC2}), we have
\bea (-1)^C=(-1)^{(F-1)N_{occ}N/2}\rho_2\eea for $C_2$-invariant insulators.

The case with $n=4$ can be similarly studied. Here four generic $\mathbf{k}$'s can be grouped with the exception of the $\Gamma$ and $M$ points which do not pair with any other $\mathbf{k}$ as they are invariant under $C_4$. Additionally, $X$ and $Y$, pair between themselves to make a group of two. The Slater determinant is then given by\begin{widetext}
\bea\label{eq:temp3}|\Phi_0\rangle&=&\prod_{i\in{occ.}}\{\gamma^\dag_i(\Gamma)\gamma^\dag_i(M)(\gamma^\dag_i(X)\gamma^\dag_i(Y))\prod_{\mathbf{k}\in{BZ/4},\mathbf{k}\neq\mathbf{k}_{inv}}(\gamma^\dag_i(k_x,k_y)\gamma^\dag_i(-k_y,k_x)\gamma^\dag_i(-k_x,-k_y)\gamma^\dag_i(k_y,-k_x))\}|0\rangle\\
\nonumber&=&(-1)^{N_{occ}N/4}\det(\mathcal{B}_{C_4}(\Gamma))\det(\mathcal{B}_{C_4}(M))\det(\mathcal{B}_{C_4}(X)\mathcal{B}_{C_4}(Y))\\
\nonumber&&\prod_{\mathbf{k}\in{BZ/4},\mathbf{k}\neq\mathbf{k}_{inv}}\mathcal{B}_{C_4}(k_x,k_y)\mathcal{B}_{C_4}(-k_y,k_x)\mathcal{B}_{C_4}(-k_x,-k_y)\mathcal{B}_{C_4}(k_y,-k_x))|\Phi_0\rangle.\eea \end{widetext}
Since $C_4$ is a four-fold rotation, two consecutive $C_4$'s equal a two-fold rotation, and that four consecutive $C_4$'s equal a complete rotation. In terms of sewing matrices, these simple facts are represented by\bea &&\mathcal{B}_{C_4}(X)\mathcal{B}_{C_4}(Y)=\mathcal{B}_{C_2}(X),\\
\nonumber&&\mathcal{B}(\bk)\mathcal{B}(C_4\bk)\mathcal{B}(C_4^2\bk)\mathcal{B}(C_4^3\bk)=(-1)^{F}.\eea By substituting these relations into Eq.(\ref{eq:temp3}), we obtain \bea\label{eq:Slater_C4}\hat{C}_4|\Phi_0\rangle=(-1)^{N_{occ}[(F-1)N/4+F]}\prod_{i\in{occ.}}\xi_i(\Gamma)\xi_i(M)\zeta_i(X)|\Phi_0\rangle.\nonumber\\\eea
Combining Eq.(\ref{eq:Slater_C4}) and Eq.(\ref{eq:ChernC4}), we have
\bea i^C=(-1)^{(F-1)N_{occ}N/4}\rho_4.\eea

Finally, we discuss systems possessing $C_3$ symmetry and $C_6$ in a straightforward fashion. Here, with the exception of the three $\mathbf{k}$'s that are invariant under $C_3$: $\Gamma$, $K$, and $K'$, whose corresponding Bloch state operators $\gamma_i(\bk_{inv})$ cannot be grouped in threes in the Slater determinant for each band, all other $\gamma_i(\bk)$'s (i.e., at other $\mathbf{k}$'s) can be grouped in threes. Following the same procedure, one finds\bea\hat{C}_3|\Phi_0\rangle=&&\prod_{i\in{occ.}}\theta_i(\Gamma)\theta_i(K)\theta_i(K')\\
\nonumber&&\prod_{\mathbf{k}\in{BZ/3},\bk\neq{\bk_{inv}}}\det(\mathcal{B}(\mathbf{k})\mathcal{B}(C_3\mathbf{k})\mathcal{B}(C_3^2\mathbf{k}))|\Phi_0\rangle.\eea
Since $\det(B(\mathbf{k})B(C_3\mathbf{k})B(C_3^2\mathbf{k}))=(-1)^{F}$ because three $C_3$'s equal a complete rotation, we have\bea\label{eq:Slater_C3}\nonumber\hat{C}_3|\Phi_0\rangle=(-1)^{FN_{occ}(N-3)/3}\prod_{i\in{occ.}}\theta_i(\Gamma)\theta_i(K)\theta_i(K')|\Phi_0\rangle.\\\eea
Therefore the Chern number can be expressed in terms of $\rho_3$ as
\bea\exp(iC\frac{2\pi}{3})=(-1)^{FN_{occ}N/3}\rho_3.\eea For $C_6$ invariant insulators, similar steps (not shown here) lead to
\bea\label{eq:Slater_C6}\exp(iC\frac{\pi}{3})=(-1)^{FN_{occ}N/6}\rho_6.\eea

Up to this point we have proved Eq.(\ref{eq:ChernSlater}) for all $C_n$-invariant insulators. Physically, it shows that the Chern number modulo $n$ in a $C_n$ invariant insulator is exactly the total angular momentum of the ground state. Although the Chern number was originally defined in terms of the Berry connection of single particle states, one can generalize its definition to an interacting system using flux insertion, also known as the distorted periodic boundary conditions, if the ground state is non-degenerate. It is then straightforward to see that in a weakly interacting system, the relation between the Chern number and the total angular momentum of ground state still holds because both numbers are quantized and therefore cannot change infinitesimally  as interaction is adiabatically turned on. This relation largely simplifies the calculation of Chern number in a weakly interacting system, because only ground state information with normal boundary condition is needed to obtain the total angular momentum, or the $C_n$ eigenvalue.

\section{Two-dimensional insulators with Dihedral Point Groups}\label{sec:diheral}

\begin{figure}[!htb]
\includegraphics[width=8cm]{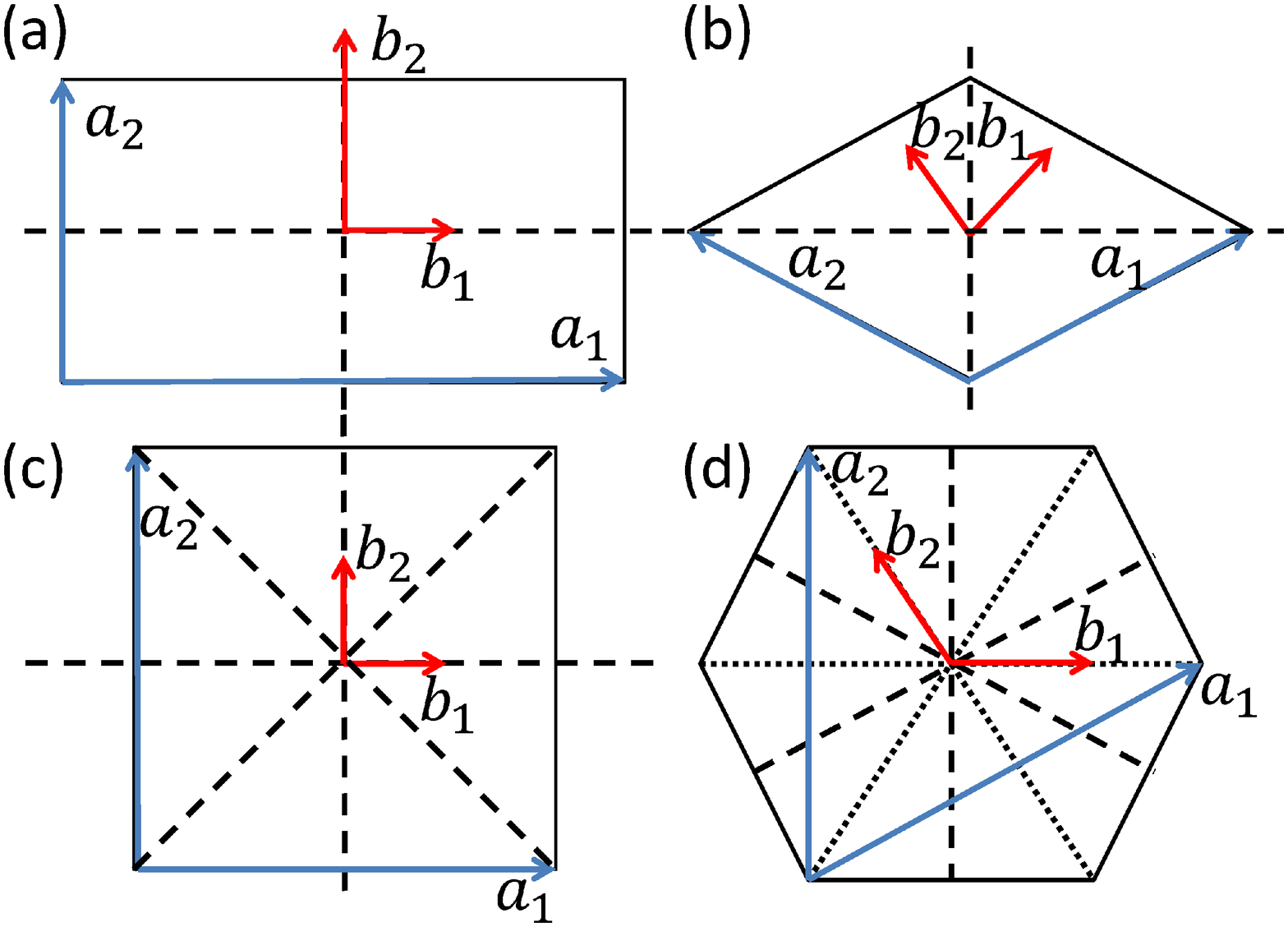}
\caption{Primitive cells of four 2D Braivis lattices: (a) rectangle lattice, (b) centered-rectangle lattice, (c) square lattice and (d) triangle lattice. The basis vectors of the real lattice, $\mathbf{a}_{1,2}$, are plotted as blue arrows and the basis vectors of the reciprocal lattice, $\mathbf{b}_{1,2}$, are plotted as red arrows. All possible $C'_2$ axes are plotted as dashed lines in (a-c). In (d), dashed lines and dotted lines represent two sets of $C'_2$ axes. $D_3$ PGS includes either one of the two sets and $D_6$ PGS contains both sets.}
\label{fig:fourBL}
\end{figure}

Out of 9 nontrivial PGS in 2D, five are called diheral PGS, denoted by $D_{n=1,2,3,4,6}$. The structure of the diheral group is simple: $D_n=C_n\cup(C'_2*C_n)$, where $C'_2$ is a two-fold rotation about some in-plane axis (referred to as $x$-axis in later text).  For later discussion, we define the basis vectors in the real space and reciprocal space in four out of five 2D Braivis lattices in Fig.\ref{fig:fourBL}: rectangular, centered rectangular, square and triangular lattices. The parallelogram lattice is excluded from the discussion as it cannot be $D_n$-invariant.

In 2D, $D_n$-group is identical to $C_{nv}$-group for spinless particles, where $C_{nv}$ is generated by $C_n$ and a mirror plane of $xz$, $M_{xz}$. This is because both $C'_2$ and $M_{xz}$ will send $(x,y)$ to $(x,-y)$. For spin-$1/2$ particles, the $C'_2$ acts as $\sigma_z$ in the spin space, while $M_{xz}$ acts as $\sigma_y$ in the spin space. Therefore, in principle a $C_{nv}$-symmetric 2D system is \emph{not} a $D_n$-symmetric system. However, throughout the section below, they are completely equivalent, because we only need a symmetry that sends a state at $(k_x,k_y)$ to one at $(k_x,-k_y)$ of the same energy and both $C'_2$ and $M_{xz}$ have this property. Therefore, all conclusions obtained here apply to $C_{nv}$-symmetric insulators without any adaptation.

\subsection{Vanishing Chern number}

A dihedral PGS has a strong constraint on the Hall conductance of a 2D insulator:
\bea\label{eq:zeroHall}\sigma_{xy}=0.\eea
Since in a noninteracting insulator the Hall conductance is proportional to the Chern number, Eq.(\ref{eq:zeroHall}) implies that the Chern number of an insulator with any dihedral PGS must be trivial. A heuristic understanding of Eq.(\ref{eq:zeroHall}) is that since $\sigma_{xy}$ changes sign under $C'_{2}$, if the system is $C'_{2}$ invariant, $\sigma_{xy}$ can only be zero. This simple argument can be put in a rigorous form presented in Appendix \ref{apndx:hall}. An alternative proof of the vanishing Chern number is by using the general transform property of field strength $\mathcal{F}_{xy}$ Eq.(\ref{eq:F_transform}) proved in Appendix \ref{apndx:transformF}, and substituting $R=C'_{2}$ to obtain
\bea\nonumber\mathcal{F}_{xy}(k_x,-k_y)=-\mathcal{B}_{C'_{2}}(k_x,k_y)\mathcal{F}_{xy}(k_x,k_y)\mathcal{B}^\dag_{C'_{2}}(k_x,k_y).\\
\label{eq:FunderC2x}\eea
Substituting Eq.(\ref{eq:FunderC2x}) into Eq.(\ref{eq:defChern}), we find
\bea \nonumber C&=&\frac{1}{2\pi}\int_{-\pi}^{\pi}dk_x[\int_{-\pi}^0Tr(\mathcal{F}_{xy}(\bk))dk_y+\int_0^{\pi}Tr(\mathcal{F}_{xy}(\bk)dk_y)\\
\nonumber&=&\frac{1}{2\pi}\int_{-\pi}^{\pi}dk_x\int_0^{\pi}Tr(\mathcal{F}_{xy}(k_x,k_y)+\mathcal{F}_{xy}(k_x,-k_y))dk_y\\
\label{eq:zeroChern} &=&0.\eea Although Eq.(\ref{eq:zeroHall}) and Eq.(\ref{eq:zeroChern}) are equivalent to each other in a noninteracting insulator, in an interacting insulator with diheral PGS, Eq.(\ref{eq:zeroChern}) loses its meaning due to the absence of Bloch states, but Eq.(\ref{eq:zeroHall}) still holds.

\subsection{Quantization of the electric polarization}

As the Chern number always vanishes in these insulators, we need some other bulk topological invariant to identify nontrivial $D_n$-invariant 2D insulators. When Chern number vanishes, one can always find a continuous gauge, or a choice of continuous Bloch wavefunctions on the whole BZ with periodic boundary condition. With this gauge choice it can be proved that the electric polarization, or the center of charge, is unambiguously defined within a unit cell, and its position vector is given by\bea\mathbf{P}=p_1\mathbf{a}_1+p_2\mathbf{a}_2,\eea where $\mathbf{a}_{1,2}$ are the unit cell basis vectors of the lattice and $p_{1,2}$ are in general within the range $[0,1)$., given by the integral of the Berry connection\cite{Resta:1994}:
\bea\label{eq:def_polarization}p_i=\frac{1}{2\pi}\int_0^1dk_1\int_0^1dk_2Tr[\mathcal{A}_i(k_1\mathbf{b}_1+k_2\mathbf{b}_2)],\eea where $k_i$ is the component of $\bk$ in its linear expansion in reciprocal lattice vectors $\mathbf{b}_{1,2}$. Here we briefly discuss why the vanishing of Chern number is necessary for the definition of polarization, taking $i=1$ for example. First notice that $\int{dk_1}Tr[\mathcal{A}_1(k_1,k_2)]=P_1(k_2)$ is the 1D polarization at fixed $k_2$. If the Chern number is nonzero, this number changes by exactly $C$ from $k_2=0$ to $k_2=1$. Therefore the integral in Eq.(\ref{eq:def_polarization}) depends on the integration range of $k_2$. For example, the integral takes different values for integrating in $[\delta,1+\delta)$ and in $[0,1)$. Since no physical quantity should depend on the choice of BZ, the polarization in this case is meaningless.

\begin{table*}
\includegraphics[width=12cm]{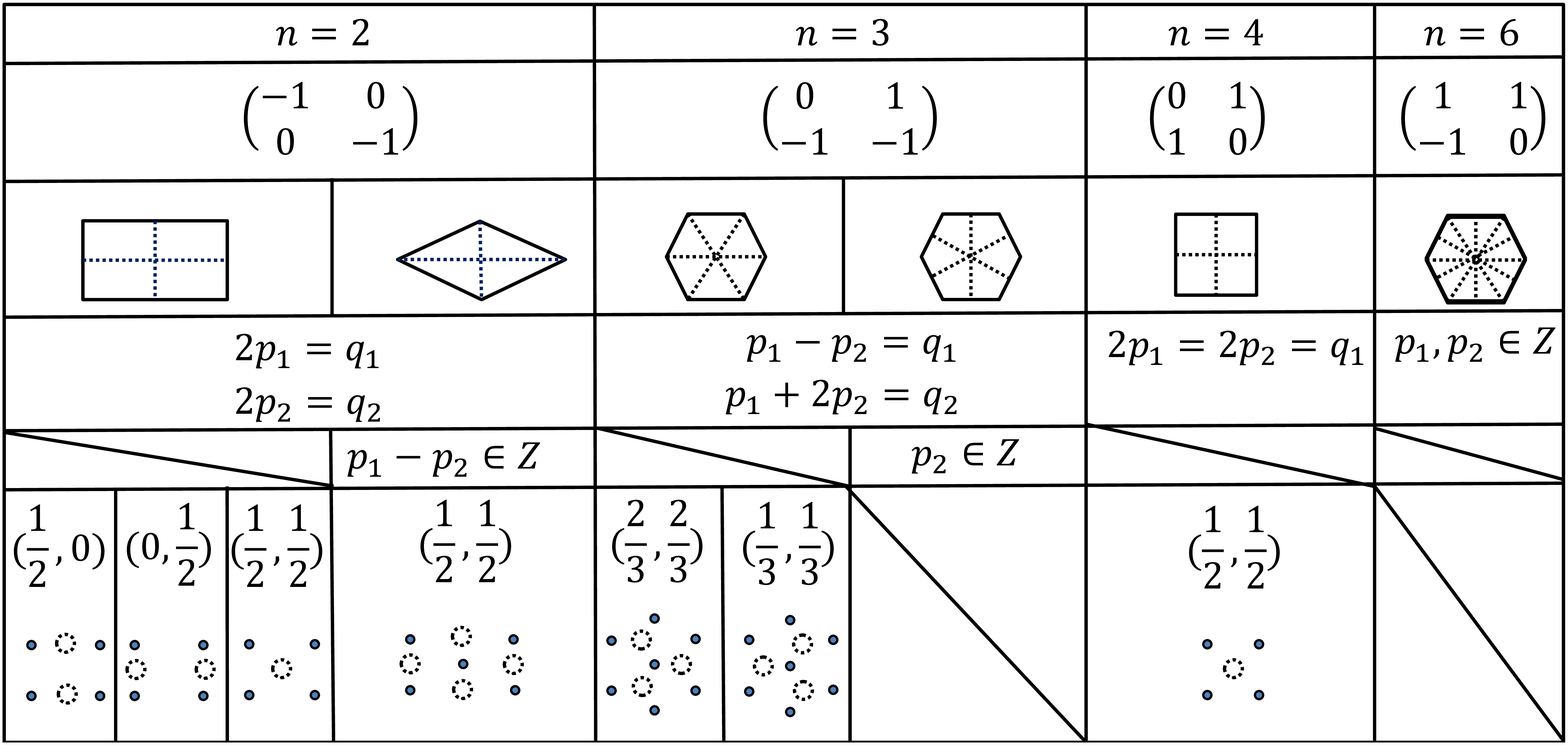}
\caption{Quantized values of electric polarization $(p_1,p_2)$ in $D_{n=2,3,4,6}$-invariant 2D insulators. From the first row: the value of $n$ in $D_n$, the transform matrix $R^{(n)}$ of $n$-fold rotation in the basis of $\mathbf{b}_{1,2}$, the unit cell in the Braivis lattice with all $C'_2$-axes, symmetry constraints on the values of $p_{1,2}$ by $C_n$, additional constraints by $C'_2$, all nontrivial $(p_1,p_2)$ pairs, the visualization of the charge center (large dotted circle) in the lattice defined by solid dots.}
\label{tab:Dn}
\end{table*}

Before presenting the formal results of quantized electric polarization in $D_n$-invariant insulators, it is helpful to establish an intuitive understanding. First we remember that the electric polarization is quantized to $0$ and $1/2$ (in unit $eL$ where $L$ is the length of the system) in a 1D insulator with inversion symmetry. This result can be easily understood in a heuristic way: suppose the system has polarization $p$, but due to inversion, the polarization is also $-p$. Then since $p$ is only well defined up to an integer, the only possible values of $p$ are $p=0$ and $p=1/2$. Now given a $D_4$-invariant system with polarization $(p_1,p_2)$, then from $C_4$-invariance it must be equivalent to $(-p_2,p_1)$, and since both $p_1$ and $p_2$ are defined up to an integer, we have equations $p_1=-p_2+integer$ and $p_2=p_1+integer$, the only solutions to which are $p_1=p_2=0$ and $p_1=p_2=1/2$. One is reminded, however, that although we only used $C_4$ to prove the quantization of $p_{1,2}$, we implicitly assumed that the Chern number is zero, without which the polarization is undefined. (This important fact is obscured in this heuristic picture.) Following the same approach, one can prove that for $D_{3,6}$-invariant insulators, $(p_1,p_2)$ is also quantized and the quantized values can be similarly written down. Despite being physically clear, this simple argument does not serve as a proof of the our statement, and, further more, it does not give the actual value of $(p_1,p_2)$ for a given band structure. In what follows, we prove the quantization of electric polarization in $D_n$-invariant insulators analytically and express the quantized polarization in terms of $C_n$-eigenvalues at points of symmetry.

Now we prove that $C_n$ PGS, a subgroup of $D_n$ PGS, quantizes $p_{1,2}$. Using sewing matrix $\mathcal{B}_{C_n}$, we have
\bea\label{eq:pitemp}p_i&=&\frac{i}{2\pi}\int{d}k_1dk_2Tr\langle{u}(\bk)|\tilde{C}_n^\dag\frac{d}{dk_i}\tilde{C}_n|u(\bk)\rangle\\
\nonumber&=&\frac{i}{2\pi}\int{d}k_1dk_2Tr\langle{u}(C_n\bk)|\mathcal{B}^\dag_{C_n}(\bk)\frac{d}{dk_i}\mathcal{B}_{C_n}(\bk)|u(C_n\bk)\rangle\\
\nonumber&=&\frac{i}{2\pi}\int{d}k_1dk_2Tr\langle{u}(C_n\bk)|\frac{d}{dk_i}|u(C_n\bk)\rangle\\
\nonumber&&+i\int{Tr}[\mathcal{B}_{C_n}(\bk)^\dag\frac{d}{dk_i}\mathcal{B}_{C_n}(\bk)]dk_1dk_2.\eea
Using the unitarity of $\mathcal{B}_{C_n}(\bk)$, the integrand in second term of the last line can be rewritten as
\bea\label{eq:pitemp3} Tr[\mathcal{B}_{C_n}(\bk)^\dag\frac{d}{dk_i}\mathcal{B}_{C_n}(\bk)]=\frac{d}{dk_i}\ln[\det(\mathcal{B}_{C_n}(\bk))].\eea
Substituting Eq.(\ref{eq:pitemp3}) to the second integral of last line of Eq.(\ref{eq:pitemp}), we find (taking $i=1$ for example while the same proceeds for $i=2$)
\bea\nonumber&&\frac{i}{2\pi}\int{}dk_1dk_2Tr[\mathcal{B}_{C_n}(k_1,k_2)^\dag\frac{d}{dk_1}\mathcal{B}_{C_n}(k_1,k_2)]\\
&=&\frac{i}{2\pi}\int{dk_2}\{\int{dk_1}\frac{d\ln[\det(\mathcal{B}_{C_n}(k_1,k_2))]}{dk_1}\}.\eea
The integral over $k_1$ will give $2q_1\pi{i}$ where $q_1$ is the winding number of $\det[\mathcal{B}_{C_n}]$, which is a pure phase, at a fixed $k_2$. However, we have chosen a smooth gauge so $q_1(k_2)$ must be a smooth function of $k_2$, which, in the case of a discrete function, is a constant. That is to say, the winding at each $k_2$ must be the same. Therefore, we have
\bea\label{eq:pitemp4}\nonumber&&\frac{i}{2\pi}\int{}dk_1dk_2Tr[\mathcal{B}_{C_n}(k_1,k_2)^\dag\frac{d}{dk_1}\mathcal{B}_{C_n}(k_1,k_2)]\\
&=&\int{dk_2}q_1(k_2)\\
&=&q_1\in{Z}.\eea
Use Eq.(\ref{eq:pitemp4}) in the last line of Eq.(\ref{eq:pitemp}), we have
\bea p_i=\frac{i}{2\pi}\int{k}_1k_2Tr\langle{u}(C_n\bk)|\frac{d}{dk_i}|u(C_n\bk)\rangle+q_i(C_n).\label{eq:pitemp5}.\eea

The first term of the right hand side of Eq.(\ref{eq:pitemp5}) may be further simplified using
\bea\label{eq:pitemp2}&&\langle{u}(C_n\bk)|\frac{d}{dk_i}|u(C_n\bk)\rangle\\
\nonumber&=&\frac{\langle{u}(C_n\bk)|u(C_n\bk+dk_iC_n\mathbf{b_i})\rangle-1}{dk_i}\\
\nonumber&=&R^{(n)}_{ij}\langle{u}(\bk')|\frac{d}{dk'_j}|u(\bk')\rangle|_{\bk'=C_n\bk},\eea where $R^{(n)}$ is the 2-by-2 matrix describing the rotation of the basis vectors $\mathbf{b}_{1,2}$ under $C_n$, summarized in the second row of Table \ref{tab:Dn}. Inserting Eq.(\ref{eq:pitemp2}) into Eq.(\ref{eq:pitemp5}) and after some rearrangement, we obtain
\bea\label{eq:pi}p_i-\sum_jR^{(n)}_{ij}p_j=q_i(C_n)\in{Z}.\eea Eq.(\ref{eq:pi}) is the general form of symmetry constraints by $C_n$-invariance on the electric polarization.

In addition to the constraints by $C_n$ symmetry, $C'_{2}$ symmetry also puts constraints on $(p_1,p_2)$. These constraints are derived in a way very similar to the way sketched in Eq.(\ref{eq:pitemp}-\ref{eq:pitemp5}). The result is given in the same form of Eq.(\ref{eq:pi}):
\bea\label{eq:pi2}p_i-\sum_jR^{'(2)}_{ij}p_j=q_i(C_n)\in{Z}.\eea Where $R^{'(2)}$ is the transformation matrix of the basis $\mathbf{b}_{1,2}$ under the two-fold rotation $C'_2$ (also listed in Table.\ref{tab:Dn}). One can see in many cases, this additional symmetry does not give any additional constraints/quantization of $p_{1,2}$, but there are a few important exceptions. For $n=1$, i.e., when there is no rotation symmetry about $z-axis$, the $C'_2$ symmetry gives quantization to the polarization perpendicular to the rotation axis (defined as $x$-axis). This is because for this system, at each $k_x$, $H(k_x,k_y)$ is a 1D inversion symmetric insulator whose polarization is quantized, and due to continuity of the gauge, the polarization is the same for all $k_x$. Therefore, the total polarization is quantized. For $D_2$-insulators, if the $C'_2$ axis coincides with the bisect of $\mathbf{a}_1$ and $\mathbf{a}_2$, i.e., in the case of centered-rectangle lattice, the $C'_2$ rotation symmetry gives $p_1-p_2\in{Z}$ as the additional constraint. Another important case in which the $C'_2$ gives additional constraint is in $D_3$-insulators, when the $C'_2$ axis coincides with $\mathbf{a}_1$ (or $\mathbf{a}_2$). There the constraint by $C'_2$ symmetry forbids any nontrivial (non-integer) values of $p_1$ and $p_2$ to exist. Row 4\&5 of Table \ref{tab:Dn} list all constraints on $(p_1,p_2)$ for insulators with $D_{n=2,3,4,6}$-invariance on four of the five 2D Braivis lattices. Notice that for $D_3$-invariant insulators, there are two possible alignment of the $C'_{2}$ axes in the triangular lattice: either along $\mathbf{a}_{1}$ (and its $C_3$ equivalent axes) or along the bisect of $\mathbf{a}_1$ and $\mathbf{a}_2$ (and its $C_3$ equivalent axes). Applying the constraints, one can easily write down all nontrivial values of $(p_1,p_2)\neq0$, listed in the last row of Table \ref{tab:Dn}. Below each pair of $(p_1,p_2)$, the electronic charge center in the real space lattice is plotted.

A few remarks are due regarding the results shown in Table \ref{tab:Dn}. First we emphasize that all results depend on the vanishing Chern number as a necessary condition, without which the 2D polarization is undefined. Second, although $C'_{2}$-invariance is a sufficient condition for the vanishing Chern number, it is not necessary for quantization of $(p_1,p_2)$. In other words, given a 2D insulator \emph{with} zero Chern number but \emph{without} $C'_{2}$ invariance, the constraints placed by $C_n$ shown in the fourth row of Table \ref{tab:Dn} are still valid. Third, we notice that for $D_3$-invariant insulators with $C'_{2}$-axis aligned with $\mathbf{a}_{1,2}$ or $D_6$-invariant insulators, $(p_1,p_2)=(0,0)$ by symmetry, i.e., there is no nontrivial polarization for these insulators.

In Sec.\ref{sec:pgs2d}, we have shown the relation between the Chern number and symmetry eigenvalues at high-symmetry points in BZ. Can we also express the electric polarization $(p_1,p_2)$ in terms of the eigenvalues of symmetry operators in $D_n$ point group at high symmetry points? To have explicit values of $p_{1,2}$ we need to evaluate the integer $q$ on the right hand side of Eq.(\ref{eq:pi}), defined through a sewing matrix in Eq.(\ref{eq:pitemp4}). From the definition, we can see that $2\pi q_i(C_n)$ is the phase difference between the determinant of the sewing matrix at two $\bk$-points separated by $\mathbf{b}_i$. In principle, $q_i$ depends on $i$ through $\mathbf{b}_i$ and PGS. In practice, due to the presence of constraints, we will see that we only need to calculate $q_1(C_2)$ and $q_1(C_3)-q_2(C_3)$, while all other $q$'s can either be derived or are not needed.

To calculate $q_1(C_2)$, first use its definition in Eq.(\ref{eq:pitemp4}) and the periodicity of the gauge to obtain
\bea\label{eq:qC22}q_1(C_2)=\frac{-i}{2\pi}\int_{-1/2}^{1/2}{dk_1}\frac{d\ln[\det(\mathcal{B}_{C_2}(k_1,0))]}{dk_1}.\eea In Eq.(\ref{eq:qC22}) we have taken $k_2=0$ because the integral is $k_2$ independent. Then we apply the sewing matrix property Eq.(\ref{eq:prod_of_B2}) and have
\bea\mathcal{B}_{C_2}(\bk)=(-1)^F\mathcal{B}^\dag_{C_2}(-\bk),\eea which leads to
\bea\label{eq:qC2}
\nonumber\ln[\det(B_{C_2}(k_1,0))]&=&iN_{occ}F\pi-\ln[\det(B_{C_2}(-k_1,0))],\\
\frac{d\ln[\det(B_{C_2}(k_1,0))]}{dk_1}&=&\frac{d\ln[\det(B_{C_2}(-k_1,0))]}{dk_1}.
\eea
Substitute Eq.(\ref{eq:qC2}) into Eq.(\ref{eq:qC22}), and we have
\bea\label{eq:qC23}q_1(C_2)&=&\frac{-i}{\pi}\int_0^{1/2}{dk_1}\frac{d\ln[\det(\mathcal{B}_{C_2}(k_1,0))]}{dk_1}\\
\nonumber&=&\frac{-i}{\pi}\ln[\frac{\det(\mathcal{B}_{C_2}(X))}{\det(\mathcal{B}_{C_2}(\Gamma))}]\;\textrm{mod}\;2\\
\nonumber&=&\frac{-i}{\pi}\ln[\prod_{n\in{occ}}\frac{\zeta_n(X)}{\zeta_n(\Gamma)}]\;\textrm{mod}\;2.\eea Eq.(\ref{eq:qC23}) can also be put in the form\bea(-1)^{q_1(C_2)}=\prod_{n\in{occ}}\frac{\zeta_n(X)}{\zeta_n(\Gamma)}.\eea The calculation of $q_2$ follows exactly the same route only replacing $X$ with $Y$. The result is
\bea(-1)^{q_2(C_2)}=\prod_{n\in{occ}}\frac{\zeta_n(Y)}{\zeta_n(\Gamma)}.\eea We then use the constraints $2p_1=q_1(C_2)$ and $2p_2=q_2(C_2)$ for $D_2$-invariant insulators to obtain the explicit values of $p_{1,2}$:
\bea(-1)^{2p_{1}}&=&\prod_{n\in{occ}}\frac{\zeta_n(X)}{\zeta_n(\Gamma)},\\
\nonumber(-1)^{2p_{2}}&=&\prod_{n\in{occ}}\frac{\zeta_n(Y)}{\zeta_n(\Gamma)}.\eea Mark that the inversion eigenvalues at $M$ do not enter the formula due to the constraint of vanishing Chern number: $\prod_{i\in{occ}}\zeta_i(\Gamma)\zeta_i(X)\zeta_i(Y)\zeta_i(M)=1$.

$q_1(C_2)$ can also be used in calculating $p_1$ in $D_4$-invariant insulators, because $D_4$ already implies $D_2$. In a $D_4$-invariant insulator, using the constraint $2p_1=2p_2=q_1(C_2)$ we find
\bea(-1)^{2p_{1,2}}=\prod_{n\in{occ}}\frac{\zeta_n(X)}{\zeta_n(\Gamma)}.\eea

\begin{figure}[!htb]
\includegraphics[width=8cm]{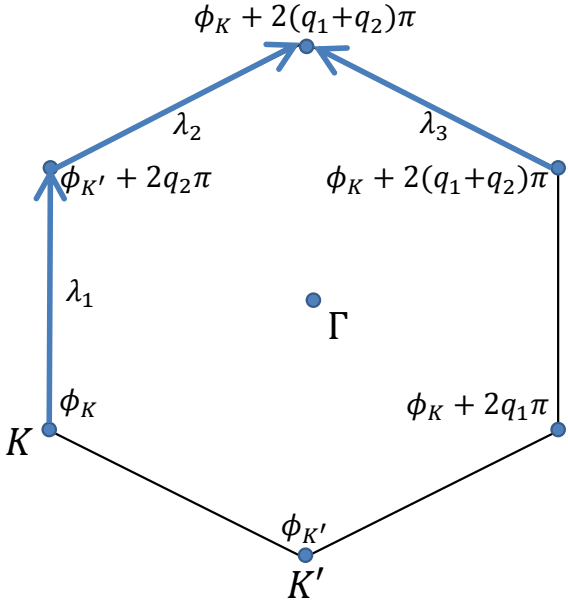}
\caption{The phase of $\det(\mathcal{B}_{C_3})$ at six corners of the BZ. $\lambda_{1,2,3}$ denote the paths of integration considered in the text.}
\label{fig:qC3}
\end{figure}

To calculate $q_1(C_3)$, we start from inspecting the phase of the determinant of the sewing matrix associated with $C_3$: \bea\phi(\bk)=-i\ln\det(\mathcal{B}_{C_3})(\bk)\eea in the BZ. Due to the continuous and periodic gauge choice, $\phi(\bk)$ is a continuous function satisfying
\bea\phi(\bk+\mathbf{b_i})=\phi(\bk)+2q_i(C_3)\pi.\eea $\phi(K)$ and $\phi(K')$ can be determined by the $C_3$ eigenvalues at these two points
\bea\phi(K)&=&-i\ln[\prod_{n\in{occ}}\theta_n(K)],\\
\nonumber\phi(K')&=&-i\ln[\prod_{n\in{occ}}\theta_n(K')].\eea We can then express $\phi(\bk)$ at all six corners of the BZ in terms of $\phi(K,K')$ and $q_{1,2}(C_3)$ as shown in Fig.\ref{fig:qC3}. Then we use the sewing matrix property
\bea\mathcal{B}_{C_3}(\bk)=(-1)^F\mathcal{B}_{C_3}^\dag(C_3\bk)\mathcal{B}^\dag(C_3^{-1}\bk)\eea to have
\bea\nonumber&&\phi(K')+2q_2\pi-\phi(K)=\int_{\lambda_1}Tr(\mathcal{B}_{C_3}(\bk)\partial\mathcal{B}_{C_3}(\bk))\cdot{d}\bk\\
\nonumber&=&(\int_{\lambda_2}{d}\bk+\int_{\lambda_3}{d}\bk)\cdot{}Tr(\mathcal{B}_{C_3}(\bk)\partial\mathcal{B}_{C_3}(\bk))\\
\label{eq:q1C3}&=&2q_1\pi-2(\phi(K')-\phi(K)),\eea where $\lambda_{1,2,3}$ are integration paths marked in Fig.\ref{fig:qC3}. From Eq.(\ref{eq:q1C3}) we extract
\bea\label{eq:q1C32}q_1(C_3)-q_2(C_3)=\frac{3}{2\pi}(\phi(K')-\phi(K)).\eea
Eq.(\ref{eq:q1C32}) and the constraints placed by $C_3$-invariance give the final expression of $p_{1,2}$:
\bea\exp(i2\pi p_{1,2})=\prod_{n\in{occ}}\frac{\theta_n(K)}{\theta_n(K')}.\eea

Up to this point, we have formulated the theory of quantized polarization in $D_n$-invariant insulators. In fact, we emphasize that the theory applies to a wider class of insulators: $C_n$-invariant insulators with vanishing Chern number. The electric polarization serves as a complimentary quantum number when the Chern number is zero (for example, in $D_n$-invariant insulators), and just like the Chern number, the electric polarization can be expressed in terms of the $C_n$-eigenvalues of bands at points of symmetry.

\section{Point Group Symmetric Topological Insulators in 3D}
\label{sec:pgs3d}
In 3D, there exist 32 different point groups, each of which is generated by four types of operators: $n$-fold rotations $C_n$, mirror plane reflections $M$ and and rotation-reflection $S_n$. $S_n$ is a combination of an $n$-fold rotation followed by a mirror reflection about the plane perpendicular to the rotation axis, or $S_n=C_n*M_{xy}$. In the following, we explore two fundamental responses of point group symmetric topological insulators: the 3D anisotropic quantum Hall conductance and the magnetoelectric polarization.

\subsection{3D Anisotropic Quantum Hall State}
The simplest 3D point groups are the cyclic groups $C_n$ as they are comprised of only one rotation axis, which we assume to be the $z$-axis. The 3D single particle Hamiltonian $\hat{H}(k_x,k_y,k_z)$ can be constructed from a series of 2D Hamiltonians $\hat{H}_{k_x,k_y;k_z}$ where $k_z$ is a momentum parameter spanning $-\pi$ to $\pi$. For every one of these individual 2D systems, one can define a Chern number $C(k_z)$. For an insulator, there is $C(k_z)=constant$, and if $C(k_z)$ changes at some $k_z$, there must be, generically, a 3D Weyl node in the bulk\cite{hughes2010inv,XuG2011,Wan2011,Fang:2011}. For $n=2,3,4,6$, $C(k_z)$ can be calculated using Eq.(\ref{eq:ChernC2}), Eq.(\ref{eq:ChernC3}), Eq.(\ref{eq:ChernC4}), and Eq.(\ref{eq:ChernC6}) respectively. Each 2D subsystem contributes $\sigma_{xy}(k_z)=Ce^2/h$, and this allows us to easily calculate the total Hall conductance of the 3D system as \bea\sigma_{xy}^{3D}=C(k_z=0)\frac{e^2L_z}{ch}\label{eq:3DQHE},\eea where $L_z$ is the length of the sample in the out-of-plane direction and $c$ is the lattice constant along the the same out-of-plane direction. On the other hand, the other two transverse conductances $\sigma_{xz}$ and $\sigma_{yz}$ must be zero as both $\sigma_{xz}$ and $\sigma_{yz}$ transform as vectors on $xy$-plane for a rotation about $z$-axis. Therefore they cannot take a non-zero value in a rotation invariant singlet ground state. Based on the preceding argument, as soon as there is more than one rotation axis, all three components of transverse conductance are quantized to zero. This directly applies that $D_n$, $D_{nd}$, $T_h$, $T_d$, $O$ and $O_h$ results in $\sigma_{xy}=\sigma_{yz}=\sigma_{xz}=0$, because they all have at least two rotation axes.

Beyond rotational symmetries, we consider the effect of mirror plane symmetries. For $C_{nh}$ point group, where the mirror plane is perpendicular to the rotation axis ($z$-axis), we can still have a non-zero $\sigma_{xy}$ given by Eq.(\ref{eq:3DQHE}). For $C_{nv}$ ($n=2,3,4,6$, while $C_{1v}=C_{1h}$) point group, since the mirror plane is parallel to $z$-axis, $\sigma_{xy}=0$, because $\sigma_{xy}$ changes sign under the mirror symmetry while the ground state is a singlet eigenstate invariant under any operation in the point group. As a result, all components of Hall conductance vanish for $C_{nv}$-invariant insulators.

Finally we discuss the Spiegel group $S_n$ ($n=2,4,6$, while $S_3$ equals $C_{3h}$ and has been discussed). $S_2$ is the same as inversion, and we borrow the result from Ref.~\cite{hughes2010inv} which states that all three transverse conductances may be quantized to nonzero values, depending on the Chern numbers on three planes defined by $k_{i=x,y,z}=0$ respectively. $S_4$ and $S_6$ has $C_2$ and $C_3$ as subgroups, respectively, and hence have $\sigma_{xz}=\sigma_{yz}=0$. To calculate $\sigma_{xy}$, we cannot directly apply Eq.(\ref{eq:ChernC4}) and Eq.(\ref{eq:ChernC6}) on $\hat{H}(k_x,k_y;k_z)$ for a generic $k_z$, because $C_4$ or $C_6$ rotational symmetry is not a symmetry of the underlying 2D subsystem. However, the subsystems with $k_z=0$ or $k_z=\pi$ are invariant under $M_{xy}$ and, on these planes, $C_4$ or $C_6$ is restored. The restored rotational invariance allows one to calculate the Chern number on $k_z=0$ or $k_z=\pi$. The insulating condition requires that the Chern number on every $k_z$ must be the same and the total Hall conductivity, $\sigma_{xy}$, is still given by Eq.(\ref{eq:3DQHE}). In fact, the insulating condition requires\bea\label{eq:S2Insulating}\prod_{i\in{occ.}}\frac{\zeta_i(\Gamma)\zeta_i(X)\zeta_i(Y)\zeta_i(S)}{\zeta_i(Z)\zeta_i(U)\zeta_i(T)\zeta_i(R)}&=&1,\\
\prod_{i\in{occ.}}\frac{\xi_i(\Gamma)\xi_i(M)\zeta_i(X)}{\xi_i(Z)\xi_i(A)\xi_i(R)}&=&1,\\
\prod_{i\in{occ.}}\frac{\eta_i(\Gamma)\theta_i(K)\zeta_i(M)}{\eta_i(A)\theta_i(H)\zeta_i(L)}&=&1,\eea
for $n=2,4,6$ respectively \cite{hughes2010inv}. Other values of the above products signal the presence of Weyl fermions.

To summarize our findings concerning the Hall conductance in 3D insulators with point group symmetries, we note that only $C_n$ for  $n=2,3,4,$ and $6$ , $C_{nh}$ ($n=2,3,4,6$) and $S_n$ ($n=2,4,6$) can have nonzero Hall conductance quantized to a multiple of the number of layers along the third direction. 3D insulators with any other point group symmetry must have $\sigma_{xy}=\sigma_{yz}=\sigma_{xz}=0$. In particular, with the exception of $S_2$, or 3D inversion, which is compatible with nonzero $\sigma_{xz}$ and $\sigma_{yz}$, all point group symmetric topological insulators have vanishing $\sigma_{xz}$ and $\sigma_{yz}$, assuming $z$-axis to be the principle axis.

\subsection{Magnetoelectric Effect}
It is known that topological insulators with time-reversal \emph{or} space inversion symmetry in 3D posses a coefficient of the magnetoelectric term $\mathbf{E}\cdot\mathbf{B}$, $P_3$, which must be quantized to either zero or one half\cite{qi2008B,hughes2010inv,Turner:2012}. We now desire to find if $P_3$ is quantized in the presence of a general PGS in a 3D insulator. We begin our discussion by considering the magnetoelectric effect in a 3D topological insulator with an arbitrary rotation symmetry, the operator of which is denoted by $\hat{R}$. In 3D, a rotation $R$ can either be a \emph{proper} or an \emph{improper} rotation. In a proper rotation, the system is rotated about a given axis in space by a given angle; while in an improper rotation is the combination of a proper rotation and space inversion, i.e., $(x,y,z)\rightarrow(-x,-y,-z)$. Mathematically, if $R_{ij}$ is the three-by-three rotation matrix, then $\det(R)=1$ for proper rotations and $\det(R)=-1$ for improper rotations. A general operation in a point group is either a proper or an improper rotation.

Before discussing the quantization of $P_3$, readers are reminded that $P_3\mathbf{E}\cdot\mathbf{B}$ is \emph{not} the only magnetoelectric response term in the Hamiltonian, It has been shown that in general there exists a cross-gap contribution, $P^o_{ij}$, a real tensor, and the total response should be $P^o_{ij}E_iB_j+P_3\mathbf{E}\cdot\mathbf{B}$\cite{Essin2010,Malashevich2010}. In any insulators with time-reversal \emph{or} space inversion symmetry, $P^o_{ij}$ is proved to be zero, which is why it is sometimes neglected; and this quantity also vanishes in the flat band limit, where all occupied bands have the same energy and all unoccupied bands have the same energy. Therefore, unlike $P_3$, $P^o_{ij}$ is \emph{not} a quantized/topological quantity so therefore does not concern our major interest in this paper. However, we find that PGS places strong constraints on the components of $P^o_{ij}$, which is detailed in Appendix \ref{apndx:OMP}.

By Eq.(\ref{eq:defSewing2}), one can use $\mathcal{B}(\mathbf{k})$ to express $|u(\mathbf{k}^{'})\rangle$ in terms of $\hat{R}|u(\mathbf{k})\rangle$, where $\bk'=R\bk$ is $\bk$ transformed by $R$:  \bea |u_a(\mathbf{k}^{'})\rangle=\sum_{b\in{occ}}\mathcal{B}^\star_{ab}(\mathbf{k})\hat{R}|u_b(\mathbf{k})\rangle.\label{eq:sewpol}\eea
Using Eq. (\ref{eq:sewpol}), the non-Abelian Berry connection $\mathcal{A}_i(k)$ then has the following property: \bea\label{eq:A_transform_C4} (\mathcal{A}_i(\mathbf{k}^{'}))_{ab}&=&-i\langle{}u_a(\mathbf{k}^{'})|R_{ij}\partial_j|u_b(\mathbf{k}^{'})\rangle\\
\nonumber&=&-iR_{ij}\sum_{c,d\in{occ}}\mathcal{B}_{ac}(\mathbf{k})\langle{u}_c(\mathbf{k})|\hat{R}^{-1}\partial_j\mathcal{B}^\star_{bd}(\mathbf{k})\hat{R}|u_d(\mathbf{k})\rangle\\
\nonumber&=&-iR_{ij}\sum_{c,d\in{occ}}\mathcal{B}_{ac}(\mathbf{k})\langle{u}_c(\mathbf{k})|\partial_j\mathcal{B}^\star_{bd}(\mathbf{k})|u_d(\mathbf{k})\rangle\\
\nonumber&=&R_{ij}(\mathcal{B}(\mathbf{k})\mathcal{A}_j(\mathbf{k})\mathcal{B}^{-1}(\mathbf{k}))_{ab}-iR_{ij}(\mathcal{B}(\mathbf{k})\partial_j\mathcal{B}(\mathbf{k}))_{ab}.\eea This non-Abelian gauge transformation exactly takes the form of the non-Ableian transform of a gauge potential, with the exception of the presence of the prefactor $R_{ij}$. The transformation of the field strength, whose proof is included in Appendix \ref{apndx:transformF}, is\bea \mathcal{F}_{ij}(\mathbf{k}^{'})=R_{ii'}R_{jj'}\mathcal{B}(\mathbf{k})\mathcal{F}_{i'j'}(\mathbf{k})\mathcal{B}^\dag(\bk)\label{eq:F_transform}.\eea

We are now in a favorable position to understand the constraints on $P_3$ imposed by the presence of general rotation symmetry $R$. In fact, one has\begin{widetext}\bea n+P_3&=&\frac{1}{16\pi^2}\int{d^3k}\epsilon_{ijk}Tr[(\mathcal{F}_{ij}(\bk)-\frac{2}{3}i\mathcal{A}_i(\bk)\mathcal{A}_j(\bk))\mathcal{A}_k(\bk)]\\
\nonumber&=&\frac{1}{16\pi^2}\int{d^3k'}\epsilon_{ijk}Tr[(\mathcal{F}_{ij}(\bk')-\frac{2}{3}i\mathcal{A}_i(\bk')\mathcal{A}_j(\bk'))\mathcal{A}_k(\bk')]\\
\nonumber&=&\frac{1}{16\pi^2}\int{d^3k}\epsilon_{ijk}R_{ii'}R_{jj'}R_{kk'}Tr[(\mathcal{B}\mathcal{F}_{i'j'}\mathcal{B}^\dag-\frac{2i}{3}(\mathcal{B}\mathcal{A}_{i'}\mathcal{B}^\dag-i\mathcal{B}\partial_{i'}\mathcal{B}^\dag)(\mathcal{B}\mathcal{A}_{j'}\mathcal{B}^\dag-i\mathcal{B}\partial_{j'}\mathcal{B}^\dag))(\mathcal{B}\mathcal{A}_{k'}\mathcal{B}^\dag-i\mathcal{B}\partial_{k'}\mathcal{B}^\dag)],\eea\end{widetext} where $n$ is a gauge dependent degree of freedom.
To proceed, notice that\bea\epsilon_{ijk}R_{ii'}R_{jj'}R_{kk'}=\det(R)\epsilon_{i'j'k'}=\pm\epsilon_{i'j'k'}.\label{eq:epsilon_transform}\eea If $R$ is a proper rotation plus sign is taken and if $R$ is an improper minus sign is taken. Then we have\begin{widetext}\bea n+P_3&=&\frac{\pm}{16\pi^2}\int{d^3k}\epsilon_{ijk}Tr[(\mathcal{B}\mathcal{F}_{ij}B^\dag-\frac{2i}{3}(B\mathcal{A}_{i}\mathcal{B}^\dag-i\mathcal{B}\partial_{i}\mathcal{B}^\dag)(\mathcal{B}\mathcal{A}_{j}\mathcal{B}^\dag-i\mathcal{B}\partial_{j}\mathcal{B}^\dag))(\mathcal{B}\mathcal{A}_{k}\mathcal{B}^\dag-i\mathcal{B}\partial_{k}\mathcal{B}^\dag)]\\
\nonumber&=&\frac{\pm}{16\pi^2}\int{d^3k}\epsilon_{ijk}Tr[(\mathcal{F}_{ij}-\frac{2}{3}i\mathcal{A}_i\mathcal{A}_j)\mathcal{A}_k]+\frac{1}{24\pi^2}\int{d^3k}Tr[(\mathcal{B}\partial_i\mathcal{B}^\dag)(\mathcal{B}\partial_j\mathcal{B}^\dag)(\mathcal{B}\partial_k\mathcal{B}^\dag)]\\\nonumber&&-\frac{i}{8\pi^2}\int{d^3k}\epsilon_{ijk}\partial_i[Tr(\mathcal{B}\mathcal{A}_j\partial_k\mathcal{B}^\dag)]\\
\nonumber&=&\pm{}(n+P_3)\pm\frac{1}{24\pi^2}\int{d^3k}Tr[(\mathcal{B}\partial_i\mathcal{B}^\dag)(\mathcal{B}\partial_j\mathcal{B}^\dag)(\mathcal{B}\partial_k\mathcal{B}^\dag)].\eea If $R$ is improper, we have \bea\label{eq:C4P3}P_3=\frac{-1}{48\pi^2}\int{d^3k}\epsilon_{ijk}Tr[(\mathcal{B}\partial_i\mathcal{B}^\dag)(\mathcal{B}\partial_j\mathcal{B}^\dag)(\mathcal{B}\partial_k\mathcal{B}^\dag)]\;\textrm{mod}\;1,\eea\end{widetext} which is indeed one half of the winding number of $B$. However, if $R$ is proper, the two $P_3$'s cancel each other and the symmetry has no constraint on $P_3$. Additionally, it gives\bea\label{eq:vanishing_winding} \frac{1}{24\pi^2}\int{d^3k}Tr[(\mathcal{B}\partial_i\mathcal{B}^\dag)(\mathcal{B}\partial_j\mathcal{B}^\dag)(\mathcal{B}\partial_k\mathcal{B}^\dag)]=0.\eea Eq. (\ref{eq:vanishing_winding}) indicates that although one may define the winding number of the sewing matrix as a quantum number, that number is always zero.

Up to this point we have formally derived the statement that only systems with improper rotation symmetry can have quantized $P_3$. In fact there is a simple way to understand this simple result. In an insulating medium with applied electromagnetic field, any point group symmetry of the medium must be preserved if one rotates the applied field together, which means that the axion term in the Hamiltonian density $P_3\mathbf{E}\cdot\mathbf{B}$ remains invariant under some improper symmetry. On the other hand, since $\mathbf{E}\cdot\mathbf{B}$ is a \emph{pseudoscaler}, under any improper rotation there is $\mathbf{E}\cdot\mathbf{B}\rightarrow-\mathbf{E}\cdot\mathbf{B}$. Therefore we have $P_3=-P_3$. From this equation one may be tempted to obtain $P_3=0$, but is reminded that $P_3$ is only well-defined up to some integer in a compact space. Therefore $P_3=-P_3$ should be interpreted as $P_3=-P_3+integer$ or $2P_3=integer$. This is the intuitive argument leading to the same quantization of $P_3$ in insulators having some improper rotation symmetry.

It is easy to check that every point group that contains an improper rotation operation must have any of $C_s$, $S_2$, or $S_4$ as its subgroup(s), where $C_s$ is a group generated by just one mirror plane. Therefore the smallest symmetry point groups that have quantized $P_3$ are $C_s$, $S_2$ and $S_4$, where it should be noted that $S_2$ is \emph{not} a subgroup of $S_4$. But this does not tell us whether it is possible to have $P_3$ quantized to the nontrivial value of $1/2$ in the presence of $C_s$, $S_2$ and $S_4$. In Ref.\onlinecite{hughes2010inv,Turner:2012}, it was made clear that there are insulators having inversion invariance without time-reversal invariance that still have nontrivial $P_3$. In the following, we will give examples of systems having only $C_s$ or only $S_4$ that have nontrivial $P_3$. We begin with a 3D Hamiltonian
\bea \mathcal{H}(\mathbf{k})&=&\sin(k_x)\Gamma_1+\sin(k_y)\Gamma_2+\sin(k_z)\Gamma_z\\
\nonumber&+&M(\mathbf{k})\Gamma_0,\eea
where $M(\mathbf{k})=3-m-\cos(k_x)-\cos(k_y)-\cos(k_y)$, $\Gamma_0=1\otimes\tau_z$, $\Gamma_1=\sigma_z\otimes\tau_x$, $\Gamma_2=1\otimes\tau_y$ and $\Gamma_3=\sigma_y\otimes\tau_x$. This Hamiltonian has time-reversal symmetry and $O_h$ point group symmetry. One can add terms to break it down to smaller point groups. To do so, we add a magnetic field along $z$-axis and an electric field along $x$-axis:\bea\delta\mathcal{H}=B\Gamma_{35}+E\Gamma_{10}.\eea The magnetic field breaks time reversal symmetry and all rotation axes except those about the $z$-axis, while the electric field further breaks rotation symmetry about $z$-axis. The only symmetry remaining is the mirror reflection about $xy$-plane. On the other hand, $P_3$ must remain unchanged in so much as the external fields are not so strong as to close the bulk gap, because a mirror reflection is still an improper rotation and quantizes $P_3$. If we choose the parameter $0<m<3$, we obtain a 3D model that has $P_3=1/2$ but no symmetry other than a reflection about $xy$-plane, or point group $C_s$. We can also add the term:
\bea \label{eq:secterm} \delta\mathcal{H}_2(\mathbf{k})=B\Gamma_{35}+t\sin(k_x)\Gamma_{15}-t\sin(k_y)\Gamma_{25}\eea to our 3D Hamiltonian. In Eq. (\ref{eq:secterm}), the first term removes time-reversal symmetry and all rotation axes except $z$-axis while the second term breaks both $C_4$ and $M_z$ separately but preserves their combination $S_4=C_4*M_{xy}$. As before, as long as the added terms are not large enough to close the bulk gap, $P_3$ remains quantized at $n+1/2$ if $0<m<3$.

While time-reversal symmetry is not the major topic of the paper but it is still interesting to discuss how this symmetry can change the previous results. Intuitively, since time-reversal operation $T$, like inversion, sends $\mathbf{k}$ to $-\mathbf{k}$, it is similar to an improper rotation symmetry as far as $P_3$ is concerned. Generally, a symmetry operation $T*R$ is equivalent to an improper/proper rotation if $R$ is a proper/improper rotation. Therefore, $P_3$ is quantized when $TR$ is a symmetry of the system for some proper $R$. From the statement is also derived a not very obvious result: if the system is \emph{not} invariant under separate inversion ($P$) or time-reversal ($T$), but is invariant under their combined operation ($P*T$), it does \emph{not} have quantized $P_3$, as $P*T$ is proper\cite{Essin2010,Malashevich2010}.

\section{Conclusion}\label{sec:conclusion}

We study several bulk topological invariants in 2D and 3D insulators with crystallographic point group symmetries, focusing on finding (i) the constraints placed by these symmetries on known topological invariants such as the Chern number and (ii) if a PGS gives rise to new topological invariants. In 2D, we show that the Chern number of a $C_n$ invariant insulator are determined up to a multiple of $n$ by by eigenvalues of $C_m$ at high-symmetry points, where $m$ divides $n$. In $D_n$-invariant insulators, we show that the Chern number is constrained to be zero, while the electric polarization, or the center of charge position is a new topological invariant, the value of which can be determined by eigenvalues of $C_m$ at high-symmetry points, where $m$ divides $n$. In 3D, we show that only $C_n$, $C_{nh}$ and $S_n$ invariant insulators can have nonzero anisotropic 3D quantum Hall conductance, while insulators with all other point group symmetries must have zero Hall conductance in every component. We also prove that the magnetoelectric susceptibility of point group symmetric topological insulators in 3D is quantized to $0$ or $1/2$, i.e., a $Z_2$ number, if and only if the point group contains at least one improper rotation.

\begin{acknowledgments}
MJG acknowledges support from the AFOSR under grant FA9550-10-1-0459 and the ONR under grant N0014-11-1-0728 and a gift the Intel Corporation. BAB was supported by NSF CAREER DMR- 095242, ONR - N00014-11-1-0635, Darpa - N66001-11- 1-4110 and David and Lucile Packard Foundation.  MJG and BAB thank the Chinese Academy of Sciences for generous hosting; BAB thanks Microsoft Station Q for generous hosting.

Note: Upon finishing this work, we are aware of another work by R. Slager, A. Mesaros, V. Juricis and J. Zaanen\cite{Slager2012} on classification of topological insulators with space group symmetries and time-reversal symmetry.

\end{acknowledgments}

\onecolumngrid
\begin{appendix}

\section{Details in Eq.(\ref{eq:NumberFromThreePoints})\label{apndx:Eq7}}
In Eq.(\ref{eq:NumberFromThreePoints}), we showed how the determinant of a Wilson loop enclosing a quarter of the BZ ($\lambda=\Gamma{X}MY\Gamma$) can be expressed in terms of the sewing matrices, but omitted several steps in which identity operators are inserted. The full details are given here:
\bea\label{eq:C4mono}&&\det(\langle{u}_i(\Gamma)|\tilde{U}_\lambda|u_j(\Gamma)\rangle)\\
\nonumber&=&\det(\langle{u}_i(\Gamma)|\tilde{U}_{\Gamma{X}}\tilde{U}_{XM}\tilde{U}_{MY}\tilde{U}_{Y\Gamma}|u_j(\Gamma)\rangle)\\
\nonumber&=&\det(\langle{u}_i(\Gamma)|\tilde{C}_4(\tilde{C}^{-1}_4\tilde{U}_{\Gamma{X}}\tilde{C}_4)\tilde{C}^{-1}_2(\tilde{C}_4\tilde{U}_{XM}\tilde{C}^{-1}_4)\tilde{C}_4\tilde{U}_{MY}\tilde{U}_{Y\Gamma}|u_j(\Gamma)\rangle)\\
\nonumber&=&\det(\langle{u}_i(\Gamma)|\tilde{C}_4\tilde{U}_{\Gamma{Y}}\tilde{C}_2^{-1}\tilde{U}_{YM}\tilde{C}_4\tilde{U}_{MY}\tilde{U}_{Y\Gamma}|u_j(\Gamma)\rangle)\\
\nonumber&=&\det(\sum_{a,b,c,d,e,f\in{occ.}}\langle{u}_i(\Gamma)|\tilde{C}_4|u_a(\Gamma)\rangle\langle{}u_a(\Gamma)|\tilde{U}_{\Gamma{Y}}|u_b(Y)\rangle\langle{u}_b(Y)|\tilde{C}_2|u_c(Y)\rangle\langle{u}_c(Y)|\tilde{U}_{YM}|u_d(M)\rangle\\
\nonumber&&\langle{u}_d(M)|\tilde{C}_4|u_e(M)\rangle\langle{u}_e(M)|\tilde{U}_{MY}|u_f(Y)\rangle\langle{u}_f(Y)|\tilde{U}_{Y\Gamma}|u_j(\Gamma)\rangle)\\
\nonumber&=&\det(\mathcal{B}_{C_4}(\Gamma)\mathcal{U}_{\Gamma{Y}}\mathcal{B}_{C_2}^{-1}(Y)\mathcal{U}_{YM}\mathcal{B}_{C_4}(M)\mathcal{U}_{MY}\mathcal{U}_{Y\gamma})\\
\nonumber&=&\det(\mathcal{B}_{C_4}(\Gamma)\mathcal{B}_{C_2}^{-1}(Y)\mathcal{B}_{C_4}(M))\det(\mathcal{U}_{\Gamma{Y}}\mathcal{U}_{YM}\mathcal{U}_{MY}\mathcal{U}_{Y\Gamma})\\
\nonumber&=&\det(\mathcal{B}_{C_4}(\Gamma)\mathcal{B}_{C_2}^{-1}(Y)\mathcal{B}_{C_4}(M)).
\eea In the steps, we notice that we insert operators like $\sum_{a\in{occ.}}|u_a(\mathbf{k})\rangle\langle{u}_a(\mathbf{k})|$ as identity operators, which is allowed if and only if $C_4$ is a symmetry of the system and the system is fully gapped. The $C_4$-symmetry guarantees the existence of another band of equal energy at $C_4\bk$ while the insulating gap guarantees no mixing from the unoccupied bands.

\section{Proof by monodromy with $C_3$ and $C_6$ symmetries\label{apndx:mono}}

In this Appendix, we prove the Eq.(\ref{eq:ChernC3}, \ref{eq:ChernC6}) using monodromy argument.

For $C_3$ invariant systems, choose a loop shown in Fig.\ref{fig:loops}(c). From $C_3$ symmetry, we have (through Eq.(\ref{eq:F_transform})) $\mathcal{F}_{xy}(\mathbf{k})=\mathcal{F}_{xy}(R_3\mathbf{k})$ and therefore the determinant of the loop integral shown in Fig.\ref{fig:loops}(c) is\bea\det(\mathcal{U}_\lambda)=e^{i2C\pi/3}\label{eq:temp4},\eea where $(\mathcal{U}_\lambda)_{ij}=\langle{}u_i(\Gamma)|\tilde{U}_\lambda|u_j(\Gamma)\rangle$. Again using $C_3$ symmetry we notice that, using Eq.(\ref{eq:transformU}), $\tilde{C}_3^{-1}\tilde{U}_{\Gamma{K^{''}}}\tilde{C}_3=\tilde{U}_{\Gamma{K^{'}}}$, and $\tilde{C}_3\tilde{U}_{K^{''}K}\tilde{C}_3^{-1}=\tilde{U}_{K^{'}{K}}$, which leads to
\bea\tilde{U}_\lambda&=&\tilde{U}_{\Gamma{K^{''}}}\tilde{U}_{K^{''}K}\tilde{U}_{KK^{'}}\tilde{U}_{K^{'}\Gamma}\\
\nonumber&=&\tilde{U}_{\Gamma{K^{''}}}\tilde{U}_{K^{''}K}\tilde{C}_3\tilde{U}_{KK^{''}}\tilde{C}^{-1}_3\tilde{C}^{-1}_3\tilde{U}_{K^{''}\Gamma}\tilde{C}_3.\eea After inserting identity operators as done in Eq.(\ref{eq:C4mono}), the determinant $\det{\mathcal{U}_\lambda}$ simplifies as
\bea\label{eq:C3mono}\det(\mathcal{U}_\lambda)&=&\det(\mathcal{U}_{\Gamma{K^{''}}}\mathcal{U}_{K^{''}K}\mathcal{B}(K)\mathcal{U}_{KK^{''}}\mathcal{B}^{-2}(K^{''})\mathcal{U}_{K^{''}\Gamma}\mathcal{B}(\Gamma))\\
\nonumber&=&\det(\mathcal{B}(K)\mathcal{B}^{-2}(K^{'})\mathcal{B}(\Gamma)).\eea In the derivation we have noticed that $K^{'}$ and $K^{''}$ are the same point by translational symmetry. Then using $\mathcal{B}^3=(-1)^F\mathcal{I}_{N_{occ.}\times N_{occ.}}$ on Eq.(\ref{eq:C3mono}) and combining Eq.(\ref{eq:temp4}), we have\bea e^{i2C\pi/3}=\prod_{i\in{occ.}}(-1)^F\theta_i(\Gamma)\theta_i(K)\theta_i(K^{'}).\eea

For $C_6$ we consider the loop shown in Fig.\ref{fig:loops}(d). The determinant of its loop integral is\bea\det(\mathcal{U}_\lambda)=e^{iC\pi/3}.\eea Notice that $\tilde{C}_6^{-1}\tilde{U}_{\Gamma{M}}\tilde{C}_6=\tilde{U}_{\Gamma{M^{'}}}$ and $\tilde{C}_3\tilde{U}_{MK}\tilde{C}^{-1}_3=\tilde{U}_{M^{'}K}$, and we have \bea\tilde{U}_\lambda&=&\tilde{U}_{\Gamma{M}}\tilde{U}_{MK}\tilde{U}_{KM'}\tilde{U}_{M'\Gamma}\\
\nonumber&=&\tilde{U}_{\Gamma{M}}\tilde{U}_{MK}\tilde{C}_3\tilde{U}_{KM}\tilde{C}^{-1}_3\tilde{C}^{-1}_6\tilde{U}_{M\Gamma}\tilde{C}_6.\eea
Inserting identity operators, the determinant becomes
\bea\label{eq:C6mono}\det(\mathcal{U}_\lambda)&=&\det(\mathcal{U}_{\Gamma{M}}\mathcal{U}_{MK}\mathcal{B}_{C_3}(K)\mathcal{U}_{KM}\mathcal{B}^{-1}_{C_2}(M)\mathcal{U}_{M\Gamma}\mathcal{B}_{C_6}(\Gamma))\\
\nonumber&=&\det(\mathcal{B}_{C_6}(\Gamma)\mathcal{B}_{C_3}(K)\mathcal{B}^{-1}_{C_2}),\eea where we have used that $C_3*C_6=C_2$. Noticing that $\tilde{C}_2^2=(-1)^{F}$, we obtain
\bea e^{iC\pi/3}=\prod_{i\in{occ.}}(-1)^F\eta_i(\Gamma)\theta_i(K)\zeta_i(K^{'}).\eea

\section{High-symmetry points in a finite system}\label{apndx:points}

In Sec.\ref{sec:pgs2d}(B), we have derived the relation between the many-body $C_n$ eigenvalue and the $C_m$ ($m$ dividing $n$) eigenvalues at high-symmetry points in the BZ, upon the condition that all these points can be taken in our system. In a real system, whether a $\bk$-point appears on the grid depends on the total number of unit cells along each direction. For example, in a 1D system, $k=\pi$ only appears when there are even number of unit cells. In this Appendix, we exhaust the possibilities of the presence/absence of each high-symmetry point in a 2D system.

On a 2D lattice with periodic boundary, a point on the grid can be generally written as
\bea
\bk=\frac{2\pi({n}_1-1)}{N_1}\mathbf{b}_1+\frac{2\pi({n}_2-1)}{N_2}\mathbf{b}_2,
\eea
where $n_1=1,...,N_1$ and $n_2=1,...,N_2$. For each high-symmetry point, the corresponding $(n_1,n_2)$ are given by:
\bea
\nonumber\Gamma:(n_1,n_2)=(1,1),\;X:(n_1,n_2)=(N_1/2,1),\;Y:(n_1,n_2)=(1,N_2/2),\;M:(n_1,n_2)=(N_1/2,N_2/2)\;\textrm{for}\;n=2,4;\\
\nonumber K:(n_1,n_2)=(N_1/3,2N_2/3),\;K':(n_1,n_2)=(2N_1/3,N_2/3)\;\textrm{for}\;n=3,6.\eea Therefore, the parity of $N_{1,2}$ determines whether a certain high-symmetry point can exist in the system. For $n=3,4,6$, we also implicitly assume that $N_1=N_2=N_0$, because otherwise the many-body system cannot have the corresponding symmetries.

Disappearance of high-symmetry points changes the results in Sec.\ref{sec:pgs2d}(B) in two aspects. First one must remove the $C_m$ eigenvalues at those points from the formulas; second there are more $\bk$-points grouped into groups of two, three, four and six, respectively, which requires correction of the prefactors in, for example, Eq.(\ref{eq:Slater_C2},\ref{eq:Slater_C3}, \ref{eq:Slater_C6}). Below we rewrite the formulas for different combinations of parities of $N_1$ and $N_2$.

For $n=2$, if $N_1=even$ and $N_2=odd$
\bea\hat{C}_2|\Phi_0\rangle=(-1)^{\frac{(F-1)N_{occ}(N-2)}{2}}\prod_{i\in{occ}}\zeta_i(\Gamma)\zeta_i(X)|\Phi_0\rangle,\eea
if $N_1=odd$ and $N_2=even$
\bea\hat{C}_2|\Phi_0\rangle=(-1)^{\frac{(F-1)N_{occ}(N-2)}{2}}\prod_{i\in{occ}}\zeta_i(\Gamma)\zeta_i(Y)|\Phi_0\rangle,\eea
if $N_1=odd$ and $N_2=odd$
\bea\hat{C}_2|\Phi_0\rangle=(-1)^{\frac{(F-1)N_{occ}(N-1)}{2}}\prod_{i\in{occ}}\zeta_i(\Gamma)|\Phi_0\rangle.\eea

For $n=3$, if $N_0\;\textrm{mod}\;3\neq0$
\bea\hat{C}_3|\Phi_0\rangle=(-1)^{FN_{occ}(N-1)/3}\prod_{i\in{occ}}\theta_i(\Gamma)|\Phi_0\rangle.\eea

For $n=4$, if $N_0=odd$
\bea\hat{C}_4|\Phi_0\rangle=(-1)^{(F-1)N_{occ}(N-1)/4}\prod_{i\in{occ}}\xi_i(\Gamma)|\Phi_0\rangle.\eea

For $n=6$, if $N_0=even$ and $N_0\;\textrm{mod}\;3\neq0$
\bea\hat{C}_6|\Phi_0\rangle=(-1)^{(F-1)N_{occ}(N-4)/6}\prod_{i\in{occ}}\eta_i(\Gamma)\zeta_i(M)|\Phi_0\rangle,\eea
if $N_0=odd$ and $N_0\;\textrm{mod}\;3=0$
\bea\hat{C}_6|\Phi_0\rangle=(-1)^{(F-1)N_{occ}(N-3)/6}\prod_{i\in{occ}}\eta_i(\Gamma)\theta_i(K)|\Phi_0\rangle,\eea
if $N_0=odd$ and $N_0\;\textrm{mod}\;3\neq0$,
\bea\hat{C}_6|\Phi_0\rangle=(-1)^{(F-1)N_{occ}(N-1)/6}\prod_{i\in{occ}}\eta_i(\Gamma).\eea

\section{Vanishing components of Hall conductance in presence of certain point group symmetries\label{apndx:hall}}

With point group symmetries, some components of Hall conductance are necessarily zero. There are three components of Hall conductance $\sigma_{xz}$, $\sigma_{yz}$ and $\sigma_{xy}$. By definition, we have $j_a=\sigma_{ab}E_b$. Under rotation $R: r_a=R_{ab}r'_b$, we have\bea R_{ab}j'_{b}=\sigma_{ab}R_{bc}E'_c,\eea or \bea j'_a=(R^{-1}\sigma{R})_{ab}E'_b.\eea In terms of quantum operators, this means \bea\hat{R}^{-1}\hat{\sigma}_{ab}\hat{R}=R_{aa'}^{-1}\hat{\sigma}_{a'b'}R_{b'b},\eea where
\bea
\hat\sigma_{ab}\equiv\lim_{\omega\rightarrow0}\frac{1-e^{-\beta\omega}}{2\omega}\int_{-\infty}^{\infty}\hat{j}_a(t)\hat{j}_b(0)e^{i\omega{t}}dt.
\eea
Specially, if $R$ is an $n$-fold rotation about $z$-axis, we have \bea\hat{R}^{-1}\hat{\sigma}_{xz}\hat{R}&=&\cos(2\pi/n)\hat{\sigma}_{xz}-\sin(2\pi/n)\hat{\sigma}_{yz},\\
\nonumber\hat{R}^{-1}\hat{\sigma}_{yz}\hat{R}&=&\cos(2\pi/n)\hat{\sigma}_{yz}+\sin(2\pi/n)\hat{\sigma}_{xz}.\eea Now we consider an insulating system with $n$-fold symmetry, and we further assume that the ground state is \emph{non-degenerate}, we have $\hat{R}|\Phi_0\rangle=\lambda(R)|\Phi_0\rangle$, because a non-degenerate ground state must be a 1D representation of the symmetry. It can be easily proved that $\sigma_{xz}=\sigma_{yz}=0$:\bea\sigma_{xz}&=&\langle\Phi_0|\hat{\sigma}_{xz}|\Phi_0\rangle\\
\nonumber&=&\frac{1}{n}\sum_{i=0,...,n-1}\langle\Phi_0|\hat{R}^{i}\hat{R}^{-i}\hat{\sigma}_{xz}\hat{R}^i\hat{R}^{-i}|\Phi_0\rangle\\
\nonumber&=&\frac{1}{n}\sum_{i=0,...,n-1}(\cos(\frac{2\pi i}{n})\sigma_{xz}-\sin(\frac{2\pi i}{n})\sigma_{yz})\\
\nonumber&=&0,\eea in which we have used $\lambda_R^*\lambda_R=1$ and the mathematical identity $\sum_{i=0,...,n-1}\cos\frac{2\pi i}{n}=\sum_{i=0,...,n-1}\sin\frac{2\pi i}{n}=0$. Specially, we have that in a 3D system with more than one rotation axis, $\sigma_{xz}=\sigma_{yz}=\sigma_{xy}=0$; and in a 2D system with mirror symmetry, $\sigma_{xy}=0$, because the mirror symmetry can be seen as a two-fold rotation about an in-plane axis.

Mark that this proof only uses the definition of Hall conductance and that the many-body ground state is a singlet. The conclusion applies to any interacting system with a singlet ground state.

For $n=even$, there exists a simpler proof utilizing the fact that any $D_n$-symmetric 2D insulator must have zero Chern number. If the system is invariant under an even-fold rotation about $z$-axis, it must also be invariant under a twofold rotation about $z$-axis. Therefore the 2D plane in $\bk$-space defined by $k_x=0$ is a 2D insulator with at least $D_1$-symmetry, therefore $\sigma_{yz}(k_x=0)=0$. Then since the quantized $\sigma_{yz}(k_x)$ must be smooth in an insulator, it must be a constant. Hence $\sigma_{yz}(k_x)=0$ for each $k_x$ and finally, the total $\sigma_{yz}=0$.

\section{Transform of field strength $\mathcal{F}$ under point group symmetry $R$ (proving Eq.(\ref{eq:F_transform}))\label{apndx:transformF}}

\bea\label{eq:transformF}\mathcal{F}_{ij}(\bk)=\partial_i\mathcal{A}_j(\bk)-\partial_j\mathcal{A}_i(\bk)+i[\mathcal{A}_i(\bk),\mathcal{A}_j(\bk)],\eea and we look at the Abelian and non-Abelian terms separately, using Eq.(\ref{eq:A_transform_C4}):\bea &&\partial_{i}\mathcal{A}_j({\bk'})-\partial_{j}{A}_i({\bk'})\\
\nonumber&=&R_{ii'}R_{jj'}[\partial_{i'}(\mathcal{B}\mathcal{A}_{j'}\mathcal{B}^\dag-i\mathcal{B}\partial_{j'}\mathcal{B}^\dag)-\partial_{j'}(\mathcal{B}\mathcal{A}_{i'}\mathcal{B}^\dag-i\mathcal{B}\partial_{i'}\mathcal{B}^\dag)]\\
\nonumber&=&R_{ii'}R_{jj'}\mathcal{B}(\partial_{i'}\mathcal{A}_{j'}-\partial_{j'}\mathcal{A}_{i'})\mathcal{B}^\dag+R_{ii'}R_{jj'}[(\partial_{i'}\mathcal{B})\mathcal{A}_{j'}\mathcal{B}^\dag+\mathcal{B}\mathcal{A}_{j'}(\partial_{i'}\mathcal{B}^\dag)-(\partial_{j'}\mathcal{B})\mathcal{A}\mathcal{B}^\dag-\mathcal{B}\mathcal{A}_{i'}\partial_{j'}\mathcal{B}^\dag]\\
\nonumber&&-iR_{ii'}R_{jj'}(\partial_{i'}\mathcal{B}\partial_{j'}\mathcal{B}^\dag-\partial_{j'}\mathcal{B}\partial_{i'}\mathcal{B}^\dag),\eea
\bea[\mathcal{A}_i(\bk'),\mathcal{A}_j(\bk')]&=&R_{ii'}R_{jj'}[\mathcal{B}\mathcal{A}_{i'}\mathcal{B}^\dag-i\mathcal{B}\partial_{i'}\mathcal{B}^\dag,\mathcal{B}\mathcal{A}_{j'}\mathcal{B}^\dag-i\mathcal{B}\partial_{j'}\mathcal{B}^\dag]\\
\nonumber&=&R_{ii'}R_{jj'}\mathcal{B}[\mathcal{A}_{i'},\mathcal{A}_{j'}]\mathcal{B}^\dag-iR_{ii'}R_{jj'}(\mathcal{B}\mathcal{A}_{i'}\partial_{j'}\mathcal{B}^\dag+(\partial_{j'}\mathcal{B})\mathcal{A}_{i'}\mathcal{B}^\dag-iR_{ii'}R_{jj'}(\partial_{i'}\mathcal{B}\mathcal{A}_{j'}\mathcal{B}^\dag+\mathcal{B}\mathcal{A}_{j'}\partial_{i'}\mathcal{B}^\dag)\\
\nonumber&&+R_{ii'}R_{jj'}(\partial_{i'}\mathcal{B}\partial_{j'}\mathcal{B}^\dag-\partial_{j'}\mathcal{B}\partial_{i'}\mathcal{B}^\dag).\eea
From these two equations, Eq.(\ref{eq:F_transform}) directly follows. (In the above expressions the argument of $\bk$ is suppressed while that of $\bk'$ is kept explicit.)

\section{Basic properties of the sewing matrix\label{apndx:sewing_property}}

Every point group symmetry operation has a certain order, i.e., there exists integer $n(R)$ for which $R^{n(R)}=E$ or $R^{n(R)}=E'$, where $E$ and $E'$ are identity operation and a $2\pi$ rotation, respectively. The distinction between $E$ and $E'$ is necessary as we are interested in the representation in Hilbert space of a single fermion. For fermions with integer spins, the representation of both $E$ and $E'$ are $\hat{I}$, but for fermions with half-integer spins, the representation for $E$ and $E'$ are $\hat{I}$ and $-\hat{I}$ respectively. For a sewing matrix associated with a point group symmetry $R$ with order $n$, we have\bea(\prod_{s=0,...,n-1}\mathcal{B}(R^{s}\bk))_{ij}&=&\sum_{i_1,i_2,...,i_n\in{occ}}\langle{}u_i(R^n\bk)|\tilde{R}|u_{i_{n-1}}(R^{n-1}\bk)\rangle\langle{}u_{i_{n-1}}(R^{n-1}\bk)|\tilde{R}|u_{i_{n-2}}(R^{n-2}\bk)\rangle...\langle{}u_{i_2}(R\bk)|\tilde{R}|u_j(\bk)\rangle\\
\nonumber&=&\langle{}u_i(R^n\bk)|\tilde{R}^n|u_j(\bk)\rangle.\eea If $R^n=E$, then we have\bea\label{eq:prod_of_B1}\prod_{s=0,...,n-1}\mathcal{B}(R^{s}\bk)=\mathcal{I}_{N_{occ.}\times N_{occ.}}.\eea If $R^n=\bar{E}$, we have \bea\label{eq:prod_of_B2}\prod_{s=0,...,n-1}\mathcal{B}(R^{s}\bk)=(-1)^F\mathcal{I}_{N_{occ.}\times N_{occ.}}.\eea

Now we show that $\mathcal{B}(\bk)$ is also a unitary matrix:
\bea\label{eq:unitarity_B}(\mathcal{B}(\bk)\mathcal{B}^\dag(\bk))_{ij}&=&\sum_{i'\in{occ}}\langle{}u_i(R\bk)|\tilde{R}|u_{i^{'}}(\bk)\rangle\langle{}u_{i^{'}}(\bk)|\tilde{R}^{\dag}|u_j(R\bk)\rangle\\
\nonumber&=&\langle{}u_i(R\bk)|u_j(R\bk)\rangle\\
\nonumber&=&\delta_{ij}.\eea In this equation we have used that the point group operator $\tilde{R}$ is unitary. This is because all symmetry operators are either unitary or antiunitary (Wigner's theorem) and a point group operation does not include either time-reversal or charge conjugation.

Using the sewing matrix, we can represent $|u_i(R\bk)\rangle$ in terms of $|u_i(\bk)\rangle$ and the sewing matrix. To see this: first notice
\bea\tilde{R}^{-1}|u_i(R\mathbf{k})\rangle=\sum_{j\in{occ}}|u_j(\mathbf{k})\rangle\langle{u}_j(\mathbf{k})|\mathcal{R}^{-1}|u_i(R\mathbf{k}))\rangle,\eea
then multiply $\tilde{R}$ on both sides and obtain
\bea|u_i(R\bk)\rangle=\sum_{j\in{occ}}\mathcal{B}^\star_{ij}(\bk)|u_j(\bk)\rangle.\eea

Finally, we consider a sewing matrix that is associated with an antiunitary operator $\tilde{R}^{'}$. First we show that the sewing matrix is still unitary:\bea(\mathcal{B}(\bk)\mathcal{B}^{\dag}(\bk))_{ij}=\sum_{m\in{occ}}\langle{}u_i(R^{'}\bk)|\tilde{R}^{'}|u_m(\bk)\rangle(\langle{}u_j(R^{'}\bk)|\tilde{R}^{'}|u_m(\bk)\rangle)^{*}.\eea Then use the antiunitarity of $\tilde{R}^{-1}$, we have
\bea(\langle{}u_j(R^{'}\bk)|\hat{R'}u_m(\bk)\rangle)^*=\langle{}u_m(\bk)|\hat{R'}^{-1}|u_j(R'\bk)\rangle.\eea And from this we obtain $\mathcal{B}(\bk)\mathcal{B}^{\dag}(\bk)=\mathcal{I}_{N_{occ}\times N_{occ}}$.

And for an antiunitary operator, we can also express $|u_i(R'\bk)\rangle$ in terms of $|u_i(\bk)\rangle$ and the sewing matrix: first notice
\bea\hat{R'}^{-1}|u_i(R'\mathbf{k})\rangle=\sum_{j\in{occ}}|u_j(\mathbf{k})\rangle\langle{u}_j(\mathbf{k})|\hat{R'}^{-1}|u_i(R'\mathbf{k}))\rangle,\eea
then act $\hat{R'}$ on both sides and have
\bea|u_i(R\bk)\rangle&=&\sum_{j\in{occ}}(\langle{u}_j(\mathbf{k})|\hat{R'}^{-1}|u_i(R'\mathbf{k})\rangle)^\star\tilde{R}'|u_j(\bk)\rangle\\
\nonumber&=&\sum_{j\in{occ}}\langle{}u_i(R'\bk)|\hat{R'}|u_j(\bk)\rangle|u_j(\bk)\rangle\\
\nonumber&=&\sum_{j\in{occ}}\mathcal{B}_{ij}|u_j(\bk)\rangle.\eea

\section{Orbital magnetoelectric polarization and point group symmetries}
\label{apndx:OMP}
According to Ref.[\onlinecite{Essin2010}], for a generic tight-binding model the orbital part of the magnetoelectric polarization includes other terms besides the Chern Simon term, given by
\bea\label{eq:OMP}
\nonumber P^o_{ij}=\frac{e^2}{\hbar}\sum_{n\in{occ},m\in{occ}}\int_{BZ}\frac{d^3k}{(2\pi)^3}\re\{\frac{\langle{u}_n(\bk)|\partial_i\tilde{P}(\bk)|u_m(\bk)\rangle\langle{u}_m(\bk)|(\partial\tilde{H}(\bk)\times\partial\tilde{P}(\bk))_j-(\partial\tilde{P}(\bk)\times\partial\tilde{H}(\bk))_j|u_n(\bk)\rangle}{E_n(\bk)-E_m(\bk)}\}.\\
\eea
When the insulator has symmetry $R$ (proper or improper), we have
\bea\label{eq:transforms}
\tilde{R}|u_n(\bk)\rangle=\sum_{n'\in{occ}}\mathcal{B}_{n'n}|u_{n'}(R\bk)\rangle,\\
\nonumber\tilde{R}|u_m(\bk)\rangle=\sum_{m'\in{unocc}}\bar{\mathcal{B}}_{n'n}|u_{n'}(R\bk)\rangle,\\
\nonumber\tilde{R}\tilde{H}(\bk)\tilde{R}^{-1}=\tilde{H}(R\bk),\\
\nonumber\tilde{R}\tilde{P}(\bk)\tilde{R}^{-1}=\tilde{P}(R\bk),
\eea where $\bar{\mathcal{B}}(\bk)$ is the sewing matrix defined for the unoccupied subspace,
\bea
\bar{\mathcal{B}}_{mn}(\bk)=\langle{u}_m(R\bk)|\tilde{R}|u_n(\bk)\rangle,
\eea where $m\in{unocc}$. Substituting Eqs.(\ref{eq:transforms}) into Eq.(\ref{eq:OMP}), we obtain:
\bea
&&\sum_{n\in{occ},m\in{unocc}}\frac{\langle{u}_n(\bk)|\partial_i\tilde{P}(\bk)|u_m(\bk)\rangle\langle{u}_m(\bk)|(\partial\tilde{H}(\bk)\times\partial\tilde{P}(\bk))_j|u_n(\bk)\rangle}{E_n(\bk)-E_m(\bk)}\\
\nonumber&=&\sum_{n,n',n''\in{occ},m,m',m''\in{unocc}}\mathcal{B}_{nn'}(\bk)\mathcal{B}^\dag_{n''n}(\bk)\bar{\mathcal{B}}_{m'm}(\bk)\bar{\mathcal{B}}^\dag_{mm''}(\bk)\times\\
\nonumber&&\frac{\langle{u}_{n'}(R\bk)|\partial_i\tilde{P}(R\bk)|u_{m'}(R\bk)\rangle\langle{u}_{m''}(R\bk)|(\partial\tilde{H}(R\bk)\times\partial\tilde{P}(R\bk))_j|u_{n''}(R\bk)\rangle}{E_n(\bk)-E_m(\bk)}.
\eea
Then we use the fact that since $\tilde{R}$ is a symmetry, the energy eigenvalue is unchanged under $\tilde{R}$, i.e., $E_m(\bk)=E_{m'}(R\bk)$ and $E_n(\bk)=E_{n'}(R\bk)$. Therefore, the summation over $m,n$ only appears in the numerator, and using the unitarity of $\mathcal{B}(\bk)$ and $\bar{\mathcal{B}}(\bk)$, we have
\bea\label{eq:rotation}
&&\sum_{n\in{occ},m\in{unocc}}\frac{\langle{u}_n(\bk)|\partial_i\tilde{P}(\bk)|u_m(\bk)\rangle\langle{u}_m(\bk)|(\partial\tilde{H}(\bk)\times\partial\tilde{P}(\bk))_j|u_n(\bk)\rangle}{E_n(\bk)-E_m(\bk)}\\
\nonumber&=&\sum_{n'\in{occ},m'\in{unocc}}\frac{\langle{u}_{n'}(R\bk)|\partial_i\tilde{P}(R\bk)|u_{m'}(R\bk)\rangle\langle{u}_{m'}(R\bk)|(\partial\tilde{H}(R\bk)\times\partial\tilde{P}(R\bk))_j|u_{n'}(R\bk)\rangle}{E_{n'}(R\bk)-E_{m'}(R\bk)}.
\eea
Define $\partial'=\partial/\partial(R\bk)=R^{-1}\partial$, then we have
\bea\label{eq:transforms2}
\partial_i\tilde{P}&=&R_{ii'}{\partial'}_{i'}\tilde{P},\\
\nonumber(\partial\tilde{H}\times\partial\tilde{P})_j&=&\det(R)R_{jj'}(\partial'\tilde{H}\times\partial'\tilde{P})_{j'}.
\eea

Combining Eqs.(\ref{eq:transforms2}), Eqs.(\ref{eq:rotation}) and Eq.(\ref{eq:OMP}), we obtain the following simple formula
\bea\label{eq:OMPtransform}
P^o_{ij}=\det(R)R_{ii'}R_{jj'}P^o_{i'j'}.
\eea
From Eq.(\ref{eq:OMPtransform}), we can see that $P^o_{ij}$ transforms like a rank-two tensor except for a factor of $\det(R)$, which is $-1$ for improper rotations. 

From this relation, we can see that PGS in general places constraints on the components of $P^o_{ij}$. Specially, for space inversion, $R=-I_{3\times3}$ and Eq.(\ref{eq:OMPtransform}) implies that every component must vanish. For mirror reflection about the $xy$-plane, we have $R_{ij}=\delta_{ij}(1-2\delta_{3j})$, and Eq.(\ref{eq:OMPtransform}) gives $P^o_{ii}=P^o_{xy}=0$, while $P^o_{xz,zx,yz,zy}$ can take nonzero values. For improper rotation $S_4$, we have $P^o_{xz,yz,zx,zy,zz}=0$ and $P^o_{xx}=-P^o_{yy}$ and $P^o_{xy}=P^o_{yx}$.
 
\end{appendix}
\twocolumngrid

\begin{thebibliography}{41}%
\makeatletter
\providecommand \@ifxundefined [1]{%
 \@ifx{#1\undefined}
}%
\providecommand \@ifnum [1]{%
 \ifnum #1\expandafter \@firstoftwo
 \else \expandafter \@secondoftwo
 \fi
}%
\providecommand \@ifx [1]{%
 \ifx #1\expandafter \@firstoftwo
 \else \expandafter \@secondoftwo
 \fi
}%
\providecommand \natexlab [1]{#1}%
\providecommand \enquote  [1]{``#1''}%
\providecommand \bibnamefont  [1]{#1}%
\providecommand \bibfnamefont [1]{#1}%
\providecommand \citenamefont [1]{#1}%
\providecommand \href@noop [0]{\@secondoftwo}%
\providecommand \href [0]{\begingroup \@sanitize@url \@href}%
\providecommand \@href[1]{\@@startlink{#1}\@@href}%
\providecommand \@@href[1]{\endgroup#1\@@endlink}%
\providecommand \@sanitize@url [0]{\catcode `\\12\catcode `\$12\catcode
  `\&12\catcode `\#12\catcode `\^12\catcode `\_12\catcode `\%12\relax}%
\providecommand \@@startlink[1]{}%
\providecommand \@@endlink[0]{}%
\providecommand \url  [0]{\begingroup\@sanitize@url \@url }%
\providecommand \@url [1]{\endgroup\@href {#1}{\urlprefix }}%
\providecommand \urlprefix  [0]{URL }%
\providecommand \Eprint [0]{\href }%
\providecommand \doibase [0]{http://dx.doi.org/}%
\providecommand \selectlanguage [0]{\@gobble}%
\providecommand \bibinfo  [0]{\@secondoftwo}%
\providecommand \bibfield  [0]{\@secondoftwo}%
\providecommand \translation [1]{[#1]}%
\providecommand \BibitemOpen [0]{}%
\providecommand \bibitemStop [0]{}%
\providecommand \bibitemNoStop [0]{.\EOS\space}%
\providecommand \EOS [0]{\spacefactor3000\relax}%
\providecommand \BibitemShut  [1]{\csname bibitem#1\endcsname}%
\let\auto@bib@innerbib\@empty
\bibitem [{\citenamefont {Laughlin}(1983)}]{laughlin1983}%
  \BibitemOpen
  \bibfield  {author} {\bibinfo {author} {\bibfnamefont {R.~B.}\ \bibnamefont
  {Laughlin}},\ }\href {\doibase 10.1103/PhysRevLett.50.1395} {\bibfield
  {journal} {\bibinfo  {journal} {Phys. Rev. Lett.}\ }\textbf {\bibinfo
  {volume} {50}},\ \bibinfo {pages} {1395} (\bibinfo {year}
  {1983})}\BibitemShut {NoStop}%
\bibitem [{\citenamefont {Wen}(1995)}]{wen1995}%
  \BibitemOpen
  \bibfield  {author} {\bibinfo {author} {\bibfnamefont {X.~G.}\ \bibnamefont
  {Wen}},\ }\href@noop {} {\bibfield  {journal} {\bibinfo  {journal} {Adv.
  Phys.}\ }\textbf {\bibinfo {volume} {44}},\ \bibinfo {pages} {405} (\bibinfo
  {year} {1995})}\BibitemShut {NoStop}%
\bibitem [{\citenamefont {Moore}\ and\ \citenamefont
  {Read}(1991)}]{Moore1991362}%
  \BibitemOpen
  \bibfield  {author} {\bibinfo {author} {\bibfnamefont {G.}~\bibnamefont
  {Moore}}\ and\ \bibinfo {author} {\bibfnamefont {N.}~\bibnamefont {Read}},\
  }\href {\doibase DOI: 10.1016/0550-3213(91)90407-O} {\bibfield  {journal}
  {\bibinfo  {journal} {Nuclear Physics B}\ }\textbf {\bibinfo {volume}
  {360}},\ \bibinfo {pages} {362 } (\bibinfo {year} {1991})}\BibitemShut
  {NoStop}%
\bibitem [{\citenamefont {\textrm{X.L. Qi}}\ \emph {et~al.}(2006)\citenamefont
  {\textrm{X.L. Qi}}, \citenamefont {\textrm{Y.S. Wu}},\ and\ \citenamefont
  {\textrm{S.C. Zhang}}}]{qi2005}%
  \BibitemOpen
  \bibfield  {author} {\bibinfo {author} {\bibnamefont {\textrm{X.L. Qi}}},
  \bibinfo {author} {\bibnamefont {\textrm{Y.S. Wu}}}, \ and\ \bibinfo {author}
  {\bibnamefont {\textrm{S.C. Zhang}}},\ }\href@noop {} {\bibfield  {journal}
  {\bibinfo  {journal} {Phys. Rev. B}\ }\textbf {\bibinfo {volume} {74}},\
  \bibinfo {pages} {085308} (\bibinfo {year} {2006})}\BibitemShut {NoStop}%
\bibitem [{\citenamefont {Kane}\ and\ \citenamefont
  {Mele}(2005)}]{Kane:2005sf}%
  \BibitemOpen
  \bibfield  {author} {\bibinfo {author} {\bibfnamefont {C.~L.}\ \bibnamefont
  {Kane}}\ and\ \bibinfo {author} {\bibfnamefont {E.~J.}\ \bibnamefont
  {Mele}},\ }\href {http://link.aps.org/abstract/PRL/v95/e146802} {\bibfield
  {journal} {\bibinfo  {journal} {Physical Review Letters}\ }\textbf {\bibinfo
  {volume} {95}} (\bibinfo {year} {2005})}\BibitemShut {NoStop}%
\bibitem [{\citenamefont {Bernevig}\ \emph {et~al.}(2006)\citenamefont
  {Bernevig}, \citenamefont {Hughes},\ and\ \citenamefont
  {Zhang}}]{Bernevig:2006kx}%
  \BibitemOpen
  \bibfield  {author} {\bibinfo {author} {\bibfnamefont {B.~A.}\ \bibnamefont
  {Bernevig}}, \bibinfo {author} {\bibfnamefont {T.~L.}\ \bibnamefont
  {Hughes}}, \ and\ \bibinfo {author} {\bibfnamefont {S.-C.}\ \bibnamefont
  {Zhang}},\ }\href
  {http://www.sciencemag.org/cgi/content/abstract/314/5806/1757} {\bibfield
  {journal} {\bibinfo  {journal} {Science}\ }\textbf {\bibinfo {volume}
  {314}},\ \bibinfo {pages} {1757} (\bibinfo {year} {2006})}\BibitemShut
  {NoStop}%
\bibitem [{\citenamefont {Fu}\ and\ \citenamefont {Kane}(2006)}]{Fu:2006rm}%
  \BibitemOpen
  \bibfield  {author} {\bibinfo {author} {\bibfnamefont {L.}~\bibnamefont
  {Fu}}\ and\ \bibinfo {author} {\bibfnamefont {C.~L.}\ \bibnamefont {Kane}},\
  }\href {http://link.aps.org/abstract/PRB/v74/e195312} {\bibfield  {journal}
  {\bibinfo  {journal} {Physical Review B (Condensed Matter and Materials
  Physics)}\ }\textbf {\bibinfo {volume} {74}},\ \bibinfo {pages} {195312}
  (\bibinfo {year} {2006})}\BibitemShut {NoStop}%
\bibitem [{\citenamefont {Fu}\ \emph {et~al.}(2007)\citenamefont {Fu},
  \citenamefont {Kane},\ and\ \citenamefont {Mele}}]{Fu:2007fk}%
  \BibitemOpen
  \bibfield  {author} {\bibinfo {author} {\bibfnamefont {L.}~\bibnamefont
  {Fu}}, \bibinfo {author} {\bibfnamefont {C.~L.}\ \bibnamefont {Kane}}, \ and\
  \bibinfo {author} {\bibfnamefont {E.~J.}\ \bibnamefont {Mele}},\ }\href
  {http://link.aps.org/abstract/PRL/v98/e106803} {\bibfield  {journal}
  {\bibinfo  {journal} {Physical Review Letters}\ }\textbf {\bibinfo {volume}
  {98}},\ \bibinfo {pages} {106803} (\bibinfo {year} {2007})}\BibitemShut
  {NoStop}%
\bibitem [{\citenamefont {Moore}\ and\ \citenamefont
  {Balents}(2007)}]{moore2007}%
  \BibitemOpen
  \bibfield  {author} {\bibinfo {author} {\bibfnamefont {J.~E.}\ \bibnamefont
  {Moore}}\ and\ \bibinfo {author} {\bibfnamefont {L.}~\bibnamefont
  {Balents}},\ }\href {\doibase 10.1103/PhysRevB.75.121306} {\bibfield
  {journal} {\bibinfo  {journal} {Phys. Rev. B}\ }\textbf {\bibinfo {volume}
  {75}},\ \bibinfo {eid} {121306} (\bibinfo {year} {2007})}\BibitemShut
  {NoStop}%
\bibitem [{\citenamefont {K\"onig}\ \emph {et~al.}(2007)\citenamefont
  {K\"onig}, \citenamefont {Wiedmann}, \citenamefont {Br\"une}, \citenamefont
  {Roth}, \citenamefont {Buhmann}, \citenamefont {Molenkamp}, \citenamefont
  {Qi},\ and\ \citenamefont {Zhang}}]{koenig2007}%
  \BibitemOpen
  \bibfield  {author} {\bibinfo {author} {\bibfnamefont {M.}~\bibnamefont
  {K\"onig}}, \bibinfo {author} {\bibfnamefont {S.}~\bibnamefont {Wiedmann}},
  \bibinfo {author} {\bibfnamefont {C.}~\bibnamefont {Br\"une}}, \bibinfo
  {author} {\bibfnamefont {A.}~\bibnamefont {Roth}}, \bibinfo {author}
  {\bibfnamefont {H.}~\bibnamefont {Buhmann}}, \bibinfo {author} {\bibfnamefont
  {L.}~\bibnamefont {Molenkamp}}, \bibinfo {author} {\bibfnamefont {X.-L.}\
  \bibnamefont {Qi}}, \ and\ \bibinfo {author} {\bibfnamefont {S.-C.}\
  \bibnamefont {Zhang}},\ }\href@noop {} {\bibfield  {journal} {\bibinfo
  {journal} {Science}\ }\textbf {\bibinfo {volume} {318}},\ \bibinfo {pages}
  {766} (\bibinfo {year} {2007})}\BibitemShut {NoStop}%
\bibitem [{\citenamefont {Zhang}\ \emph {et~al.}(2009)\citenamefont {Zhang},
  \citenamefont {Liu}, \citenamefont {Qi}, \citenamefont {Dai}, \citenamefont
  {Fang},\ and\ \citenamefont {Zhang}}]{zhang2009}%
  \BibitemOpen
  \bibfield  {author} {\bibinfo {author} {\bibfnamefont {H.}~\bibnamefont
  {Zhang}}, \bibinfo {author} {\bibfnamefont {C.-X.}\ \bibnamefont {Liu}},
  \bibinfo {author} {\bibfnamefont {X.-L.}\ \bibnamefont {Qi}}, \bibinfo
  {author} {\bibfnamefont {X.}~\bibnamefont {Dai}}, \bibinfo {author}
  {\bibfnamefont {Z.}~\bibnamefont {Fang}}, \ and\ \bibinfo {author}
  {\bibfnamefont {S.-C.}\ \bibnamefont {Zhang}},\ }\href@noop {} {\bibfield
  {journal} {\bibinfo  {journal} {Nat. Phys.}\ }\textbf {\bibinfo {volume}
  {5}},\ \bibinfo {pages} {438} (\bibinfo {year} {2009})}\BibitemShut {NoStop}%
\bibitem [{\citenamefont {Fu}\ and\ \citenamefont {Kane}(2008)}]{fu2008}%
  \BibitemOpen
  \bibfield  {author} {\bibinfo {author} {\bibfnamefont {L.}~\bibnamefont
  {Fu}}\ and\ \bibinfo {author} {\bibfnamefont {C.~L.}\ \bibnamefont {Kane}},\
  }\href@noop {} {\bibfield  {journal} {\bibinfo  {journal} {Phys. Rev. Lett.}\
  }\textbf {\bibinfo {volume} {100}},\ \bibinfo {pages} {096407} (\bibinfo
  {year} {2008})}\BibitemShut {NoStop}%
\bibitem [{\citenamefont {Kitaev}(2009)}]{Kitaev}%
  \BibitemOpen
  \bibfield  {author} {\bibinfo {author} {\bibfnamefont {A.}~\bibnamefont
  {Kitaev}},\ }\href@noop {} {\bibfield  {journal} {\bibinfo  {journal} {AIP
  Conf. Proc.}\ }\textbf {\bibinfo {volume} {1134}},\ \bibinfo {pages}
  {arXiv:0901.2686.} (\bibinfo {year} {2009})}\BibitemShut {NoStop}%
\bibitem [{\citenamefont {Schnyder}\ \emph {et~al.}(2008)\citenamefont
  {Schnyder}, \citenamefont {Ryu}, \citenamefont {Furusaki},\ and\
  \citenamefont {Ludwig}}]{schnyder2008}%
  \BibitemOpen
  \bibfield  {author} {\bibinfo {author} {\bibfnamefont {A.~P.}\ \bibnamefont
  {Schnyder}}, \bibinfo {author} {\bibfnamefont {S.}~\bibnamefont {Ryu}},
  \bibinfo {author} {\bibfnamefont {A.}~\bibnamefont {Furusaki}}, \ and\
  \bibinfo {author} {\bibfnamefont {A.~W.~W.}\ \bibnamefont {Ludwig}},\
  }\href@noop {} {\bibfield  {journal} {\bibinfo  {journal} {Phys. Rev. B}\
  }\textbf {\bibinfo {volume} {78}},\ \bibinfo {pages} {195125} (\bibinfo
  {year} {2008})}\BibitemShut {NoStop}%
\bibitem [{\citenamefont {Sun}\ \emph {et~al.}(2011)\citenamefont {Sun},
  \citenamefont {Liu}, \citenamefont {Hemmerich},\ and\ \citenamefont
  {Das~Sarma}}]{SunK2011}%
  \BibitemOpen
  \bibfield  {author} {\bibinfo {author} {\bibfnamefont {K.}~\bibnamefont
  {Sun}}, \bibinfo {author} {\bibfnamefont {W.~V.}\ \bibnamefont {Liu}},
  \bibinfo {author} {\bibfnamefont {A.}~\bibnamefont {Hemmerich}}, \ and\
  \bibinfo {author} {\bibfnamefont {S.}~\bibnamefont {Das~Sarma}},\ }\href@noop
  {} {\bibfield  {journal} {\bibinfo  {journal} {Nature Physics}\ }\textbf
  {\bibinfo {volume} {8}},\ \bibinfo {pages} {67} (\bibinfo {year}
  {2011})}\BibitemShut {NoStop}%
\bibitem [{\citenamefont {Wan}\ \emph {et~al.}(2011)\citenamefont {Wan},
  \citenamefont {Turner}, \citenamefont {Vishwanath},\ and\ \citenamefont
  {Savrasov}}]{Wan2011}%
  \BibitemOpen
  \bibfield  {author} {\bibinfo {author} {\bibfnamefont {X.}~\bibnamefont
  {Wan}}, \bibinfo {author} {\bibfnamefont {A.}~\bibnamefont {Turner}},
  \bibinfo {author} {\bibfnamefont {A.}~\bibnamefont {Vishwanath}}, \ and\
  \bibinfo {author} {\bibfnamefont {S.~Y.}\ \bibnamefont {Savrasov}},\
  }\href@noop {} {\bibfield  {journal} {\bibinfo  {journal} {Phys. Rev. B}\
  }\textbf {\bibinfo {volume} {83}},\ \bibinfo {pages} {205101} (\bibinfo
  {year} {2011})}\BibitemShut {NoStop}%
\bibitem [{\citenamefont {Xu}\ \emph {et~al.}(2011)\citenamefont {Xu},
  \citenamefont {Weng}, \citenamefont {Wang}, \citenamefont {Dai},\ and\
  \citenamefont {Fang}}]{XuG2011}%
  \BibitemOpen
  \bibfield  {author} {\bibinfo {author} {\bibfnamefont {G.}~\bibnamefont
  {Xu}}, \bibinfo {author} {\bibfnamefont {H.}~\bibnamefont {Weng}}, \bibinfo
  {author} {\bibfnamefont {Z.}~\bibnamefont {Wang}}, \bibinfo {author}
  {\bibfnamefont {X.}~\bibnamefont {Dai}}, \ and\ \bibinfo {author}
  {\bibfnamefont {Z.}~\bibnamefont {Fang}},\ }\href@noop {} {\bibfield
  {journal} {\bibinfo  {journal} {Phys. Rev. Lett.}\ }\textbf {\bibinfo
  {volume} {107}},\ \bibinfo {pages} {186806} (\bibinfo {year}
  {2011})}\BibitemShut {NoStop}%
\bibitem [{\citenamefont {Fang}\ \emph {et~al.}(2012)\citenamefont {Fang},
  \citenamefont {Gilbert}, \citenamefont {Dai},\ and\ \citenamefont
  {Bernevig}}]{Fang:2011}%
  \BibitemOpen
  \bibfield  {author} {\bibinfo {author} {\bibfnamefont {C.}~\bibnamefont
  {Fang}}, \bibinfo {author} {\bibfnamefont {M.~J.}\ \bibnamefont {Gilbert}},
  \bibinfo {author} {\bibfnamefont {X.}~\bibnamefont {Dai}}, \ and\ \bibinfo
  {author} {\bibfnamefont {B.~A.}\ \bibnamefont {Bernevig}},\ }\href {\doibase
  10.1103/PhysRevLett.108.266802} {\bibfield  {journal} {\bibinfo  {journal}
  {Phys. Rev. Lett.}\ }\textbf {\bibinfo {volume} {108}},\ \bibinfo {pages}
  {266802} (\bibinfo {year} {2012})}\BibitemShut {NoStop}%
\bibitem [{\citenamefont {Hal\'asz}\ and\ \citenamefont
  {Balents}(2012)}]{HalaszG2012}%
  \BibitemOpen
  \bibfield  {author} {\bibinfo {author} {\bibfnamefont {G.~B.}\ \bibnamefont
  {Hal\'asz}}\ and\ \bibinfo {author} {\bibfnamefont {L.}~\bibnamefont
  {Balents}},\ }\href {\doibase 10.1103/PhysRevB.85.035103} {\bibfield
  {journal} {\bibinfo  {journal} {Phys. Rev. B}\ }\textbf {\bibinfo {volume}
  {85}},\ \bibinfo {pages} {035103} (\bibinfo {year} {2012})}\BibitemShut
  {NoStop}%
\bibitem [{\citenamefont {Klitzing}\ \emph {et~al.}(1980)\citenamefont
  {Klitzing}, \citenamefont {Dorda},\ and\ \citenamefont {Pepper}}]{Klitzing}%
  \BibitemOpen
  \bibfield  {author} {\bibinfo {author} {\bibfnamefont {K.~v.}\ \bibnamefont
  {Klitzing}}, \bibinfo {author} {\bibfnamefont {G.}~\bibnamefont {Dorda}}, \
  and\ \bibinfo {author} {\bibfnamefont {M.}~\bibnamefont {Pepper}},\ }\href
  {\doibase 10.1103/PhysRevLett.45.494} {\bibfield  {journal} {\bibinfo
  {journal} {Phys. Rev. Lett.}\ }\textbf {\bibinfo {volume} {45}},\ \bibinfo
  {pages} {494} (\bibinfo {year} {1980})}\BibitemShut {NoStop}%
\bibitem [{\citenamefont {Thouless}\ \emph {et~al.}(1982)\citenamefont
  {Thouless}, \citenamefont {Kohmoto}, \citenamefont {Nightingale},\ and\
  \citenamefont {den Nijs}}]{Thouless:1982rz}%
  \BibitemOpen
  \bibfield  {author} {\bibinfo {author} {\bibfnamefont {D.~J.}\ \bibnamefont
  {Thouless}}, \bibinfo {author} {\bibfnamefont {M.}~\bibnamefont {Kohmoto}},
  \bibinfo {author} {\bibfnamefont {M.~P.}\ \bibnamefont {Nightingale}}, \ and\
  \bibinfo {author} {\bibfnamefont {M.}~\bibnamefont {den Nijs}},\ }\href
  {http://link.aps.org/abstract/PRL/v49/p405} {\bibfield  {journal} {\bibinfo
  {journal} {Physical Review Letters}\ }\textbf {\bibinfo {volume} {49}}
  (\bibinfo {year} {1982})}\BibitemShut {NoStop}%
\bibitem [{\citenamefont {Haldane}(1988)}]{Haldane1988}%
  \BibitemOpen
  \bibfield  {author} {\bibinfo {author} {\bibfnamefont {F.~D.~M.}\
  \bibnamefont {Haldane}},\ }\href {http://link.aps.org/abstract/PRL/v61/p2015}
  {\bibfield  {journal} {\bibinfo  {journal} {Physical Review Letters}\
  }\textbf {\bibinfo {volume} {61}} (\bibinfo {year} {1988})}\BibitemShut
  {NoStop}%
\bibitem [{()}]{}%
  \BibitemOpen
  \href@noop {} {\ }\BibitemShut {NoStop}%
\bibitem [{\citenamefont {Qi}\ \emph {et~al.}(2008{\natexlab{a}})\citenamefont
  {Qi}, \citenamefont {Hughes},\ and\ \citenamefont {Zhang}}]{qi2008B}%
  \BibitemOpen
  \bibfield  {author} {\bibinfo {author} {\bibfnamefont {X.-L.}\ \bibnamefont
  {Qi}}, \bibinfo {author} {\bibfnamefont {T.}~\bibnamefont {Hughes}}, \ and\
  \bibinfo {author} {\bibfnamefont {S.-C.}\ \bibnamefont {Zhang}},\ }\href@noop
  {} {\bibfield  {journal} {\bibinfo  {journal} {Phys. Rev. B}\ }\textbf
  {\bibinfo {volume} {78}},\ \bibinfo {pages} {195424} (\bibinfo {year}
  {2008}{\natexlab{a}})}\BibitemShut {NoStop}%
\bibitem [{\citenamefont {Qi}\ and\ \citenamefont {Zhang}(2011)}]{qi2011rev}%
  \BibitemOpen
  \bibfield  {author} {\bibinfo {author} {\bibfnamefont {X.~L.}\ \bibnamefont
  {Qi}}\ and\ \bibinfo {author} {\bibfnamefont {S.~C.}\ \bibnamefont {Zhang}},\
  }\href@noop {} {\bibfield  {journal} {\bibinfo  {journal} {Rev. Mod. Phys.}\
  }\textbf {\bibinfo {volume} {83}},\ \bibinfo {pages} {1057} (\bibinfo {year}
  {2011})}\BibitemShut {NoStop}%
\bibitem [{\citenamefont {Qi}\ \emph {et~al.}(2008{\natexlab{b}})\citenamefont
  {Qi}, \citenamefont {Hughes},\ and\ \citenamefont {Zhang}}]{Qi:2008sf}%
  \BibitemOpen
  \bibfield  {author} {\bibinfo {author} {\bibfnamefont {X.-L.}\ \bibnamefont
  {Qi}}, \bibinfo {author} {\bibfnamefont {T.~L.}\ \bibnamefont {Hughes}}, \
  and\ \bibinfo {author} {\bibfnamefont {S.-C.}\ \bibnamefont {Zhang}},\ }\href
  {http://link.aps.org/abstract/PRB/v78/e195424} {\bibfield  {journal}
  {\bibinfo  {journal} {Physical Review B (Condensed Matter and Materials
  Physics)}\ }\textbf {\bibinfo {volume} {78}},\ \bibinfo {pages} {195424}
  (\bibinfo {year} {2008}{\natexlab{b}})}\BibitemShut {NoStop}%
\bibitem [{\citenamefont {Hsieh}\ \emph {et~al.}(2008)\citenamefont {Hsieh},
  \citenamefont {Qian}, \citenamefont {Wray}, \citenamefont {Xia},
  \citenamefont {Hor}, \citenamefont {Cava},\ and\ \citenamefont
  {Hasan}}]{Hsieh:2008fk}%
  \BibitemOpen
  \bibfield  {author} {\bibinfo {author} {\bibfnamefont {D.}~\bibnamefont
  {Hsieh}}, \bibinfo {author} {\bibfnamefont {D.}~\bibnamefont {Qian}},
  \bibinfo {author} {\bibfnamefont {L.}~\bibnamefont {Wray}}, \bibinfo {author}
  {\bibfnamefont {Y.}~\bibnamefont {Xia}}, \bibinfo {author} {\bibfnamefont
  {Y.~S.}\ \bibnamefont {Hor}}, \bibinfo {author} {\bibfnamefont {R.~J.}\
  \bibnamefont {Cava}}, \ and\ \bibinfo {author} {\bibfnamefont {M.~Z.}\
  \bibnamefont {Hasan}},\ }\href {http://dx.doi.org/10.1038/nature06843}
  {\bibfield  {journal} {\bibinfo  {journal} {Nature}\ }\textbf {\bibinfo
  {volume} {452}},\ \bibinfo {pages} {970} (\bibinfo {year}
  {2008})}\BibitemShut {NoStop}%
\bibitem [{\citenamefont {Hsieh}\ \emph
  {et~al.}(2009{\natexlab{a}})\citenamefont {Hsieh}, \citenamefont {Xia},
  \citenamefont {Qian}, \citenamefont {Wray}, \citenamefont {Dil},
  \citenamefont {Meier}, \citenamefont {Patthey}, \citenamefont {Osterwalder},
  \citenamefont {Fedorov}, \citenamefont {H.~Lin}, \citenamefont {Grauer},
  \citenamefont {Hor}, \citenamefont {Cava},\ and\ \citenamefont
  {Hasan}}]{hsieh2009a}%
  \BibitemOpen
  \bibfield  {author} {\bibinfo {author} {\bibfnamefont {D.}~\bibnamefont
  {Hsieh}}, \bibinfo {author} {\bibfnamefont {Y.}~\bibnamefont {Xia}}, \bibinfo
  {author} {\bibfnamefont {D.}~\bibnamefont {Qian}}, \bibinfo {author}
  {\bibfnamefont {L.}~\bibnamefont {Wray}}, \bibinfo {author} {\bibfnamefont
  {J.~H.}\ \bibnamefont {Dil}}, \bibinfo {author} {\bibfnamefont
  {F.}~\bibnamefont {Meier}}, \bibinfo {author} {\bibfnamefont
  {L.}~\bibnamefont {Patthey}}, \bibinfo {author} {\bibfnamefont
  {J.}~\bibnamefont {Osterwalder}}, \bibinfo {author} {\bibfnamefont
  {A.}~\bibnamefont {Fedorov}}, \bibinfo {author} {\bibfnamefont {A.~B.}\
  \bibnamefont {H.~Lin}}, \bibinfo {author} {\bibfnamefont {D.}~\bibnamefont
  {Grauer}}, \bibinfo {author} {\bibfnamefont {Y.}~\bibnamefont {Hor}},
  \bibinfo {author} {\bibfnamefont {R.}~\bibnamefont {Cava}}, \ and\ \bibinfo
  {author} {\bibfnamefont {M.}~\bibnamefont {Hasan}},\ }\href@noop {}
  {\bibfield  {journal} {\bibinfo  {journal} {Nature}\ }\textbf {\bibinfo
  {volume} {460}},\ \bibinfo {pages} {1101} (\bibinfo {year}
  {2009}{\natexlab{a}})}\BibitemShut {NoStop}%
\bibitem [{\citenamefont {Hsieh}\ \emph
  {et~al.}(2009{\natexlab{b}})\citenamefont {Hsieh}, \citenamefont {Xia},
  \citenamefont {Wray}, \citenamefont {Qian}, \citenamefont {Pal},
  \citenamefont {Dil}, \citenamefont {Osterwalder}, \citenamefont {Meier},
  \citenamefont {Bihlmayer}, \citenamefont {Kane}, \citenamefont {Hor},
  \citenamefont {Cava},\ and\ \citenamefont {Hasan}}]{Hsieh2009}%
  \BibitemOpen
  \bibfield  {author} {\bibinfo {author} {\bibfnamefont {D.}~\bibnamefont
  {Hsieh}}, \bibinfo {author} {\bibfnamefont {Y.}~\bibnamefont {Xia}}, \bibinfo
  {author} {\bibfnamefont {L.}~\bibnamefont {Wray}}, \bibinfo {author}
  {\bibfnamefont {D.}~\bibnamefont {Qian}}, \bibinfo {author} {\bibfnamefont
  {A.}~\bibnamefont {Pal}}, \bibinfo {author} {\bibfnamefont {J.~H.}\
  \bibnamefont {Dil}}, \bibinfo {author} {\bibfnamefont {J.}~\bibnamefont
  {Osterwalder}}, \bibinfo {author} {\bibfnamefont {F.}~\bibnamefont {Meier}},
  \bibinfo {author} {\bibfnamefont {G.}~\bibnamefont {Bihlmayer}}, \bibinfo
  {author} {\bibfnamefont {C.~L.}\ \bibnamefont {Kane}}, \bibinfo {author}
  {\bibfnamefont {Y.~S.}\ \bibnamefont {Hor}}, \bibinfo {author} {\bibfnamefont
  {R.~J.}\ \bibnamefont {Cava}}, \ and\ \bibinfo {author} {\bibfnamefont
  {M.~Z.}\ \bibnamefont {Hasan}},\ }\href@noop {} {\bibfield  {journal}
  {\bibinfo  {journal} {Science}\ }\textbf {\bibinfo {volume} {323}},\ \bibinfo
  {pages} {919} (\bibinfo {year} {2009}{\natexlab{b}})}\BibitemShut {NoStop}%
\bibitem [{\citenamefont {Liu}\ \emph {et~al.}(2009)\citenamefont {Liu},
  \citenamefont {Liu}, \citenamefont {Xu}, \citenamefont {Qi},\ and\
  \citenamefont {Zhang}}]{liu2009}%
  \BibitemOpen
  \bibfield  {author} {\bibinfo {author} {\bibfnamefont {Q.}~\bibnamefont
  {Liu}}, \bibinfo {author} {\bibfnamefont {C.-X.}\ \bibnamefont {Liu}},
  \bibinfo {author} {\bibfnamefont {C.}~\bibnamefont {Xu}}, \bibinfo {author}
  {\bibfnamefont {X.-L.}\ \bibnamefont {Qi}}, \ and\ \bibinfo {author}
  {\bibfnamefont {S.-C.}\ \bibnamefont {Zhang}},\ }\href@noop {} {\bibfield
  {journal} {\bibinfo  {journal} {Phys. Rev. Lett.}\ }\textbf {\bibinfo
  {volume} {102}},\ \bibinfo {pages} {156603} (\bibinfo {year}
  {2009})}\BibitemShut {NoStop}%
\bibitem [{\citenamefont {Zhou}\ \emph {et~al.}(2009)\citenamefont {Zhou},
  \citenamefont {Fang}, \citenamefont {Tsai},\ and\ \citenamefont
  {Hu}}]{ZhouX2009}%
  \BibitemOpen
  \bibfield  {author} {\bibinfo {author} {\bibfnamefont {X.}~\bibnamefont
  {Zhou}}, \bibinfo {author} {\bibfnamefont {C.}~\bibnamefont {Fang}}, \bibinfo
  {author} {\bibfnamefont {W.-F.}\ \bibnamefont {Tsai}}, \ and\ \bibinfo
  {author} {\bibfnamefont {J.}~\bibnamefont {Hu}},\ }\href@noop {} {\bibfield
  {journal} {\bibinfo  {journal} {Phys. Rev. B}\ }\textbf {\bibinfo {volume}
  {80}},\ \bibinfo {pages} {245317} (\bibinfo {year} {2009})}\BibitemShut
  {NoStop}%
\bibitem [{\citenamefont {Fu}(2011)}]{Fu:2011}%
  \BibitemOpen
  \bibfield  {author} {\bibinfo {author} {\bibfnamefont {L.}~\bibnamefont
  {Fu}},\ }\href@noop {} {\bibfield  {journal} {\bibinfo  {journal} {Phys. Rev.
  Lett.}\ }\textbf {\bibinfo {volume} {106}},\ \bibinfo {pages} {106802}
  (\bibinfo {year} {2011})}\BibitemShut {NoStop}%
\bibitem [{\citenamefont {Hughes}\ \emph {et~al.}(2010)\citenamefont {Hughes},
  \citenamefont {Prodan},\ and\ \citenamefont {Bernevig}}]{hughes2010inv}%
  \BibitemOpen
  \bibfield  {author} {\bibinfo {author} {\bibfnamefont {T.~L.}\ \bibnamefont
  {Hughes}}, \bibinfo {author} {\bibfnamefont {E.}~\bibnamefont {Prodan}}, \
  and\ \bibinfo {author} {\bibfnamefont {B.~A.}\ \bibnamefont {Bernevig}},\
  }\href@noop {} {\bibfield  {journal} {\bibinfo  {journal} {Phys. Rev. B}\
  }\textbf {\bibinfo {volume} {83}},\ \bibinfo {pages} {245132} (\bibinfo
  {year} {2010})}\BibitemShut {NoStop}%
\bibitem [{\citenamefont {Turner}\ \emph {et~al.}(2012)\citenamefont {Turner},
  \citenamefont {Zhang}, \citenamefont {Mong},\ and\ \citenamefont
  {Vishwanath}}]{Turner:2012}%
  \BibitemOpen
  \bibfield  {author} {\bibinfo {author} {\bibfnamefont {A.~M.}\ \bibnamefont
  {Turner}}, \bibinfo {author} {\bibfnamefont {Y.}~\bibnamefont {Zhang}},
  \bibinfo {author} {\bibfnamefont {R.~S.~K.}\ \bibnamefont {Mong}}, \ and\
  \bibinfo {author} {\bibfnamefont {A.}~\bibnamefont {Vishwanath}},\ }\href
  {\doibase 10.1103/PhysRevB.85.165120} {\bibfield  {journal} {\bibinfo
  {journal} {Phys. Rev. B}\ }\textbf {\bibinfo {volume} {85}},\ \bibinfo
  {pages} {165120} (\bibinfo {year} {2012})}\BibitemShut {NoStop}%
\bibitem [{\citenamefont {Fu}\ and\ \citenamefont {Kane}(2007)}]{fu2007a}%
  \BibitemOpen
  \bibfield  {author} {\bibinfo {author} {\bibfnamefont {L.}~\bibnamefont
  {Fu}}\ and\ \bibinfo {author} {\bibfnamefont {C.~L.}\ \bibnamefont {Kane}},\
  }\href {\doibase 10.1103/PhysRevB.76.045302} {\bibfield  {journal} {\bibinfo
  {journal} {Phys. Rev. B}\ }\textbf {\bibinfo {volume} {76}},\ \bibinfo {eid}
  {045302} (\bibinfo {year} {2007})}\BibitemShut {NoStop}%
\bibitem [{\citenamefont {Hsieh}\ \emph {et~al.}(2012)\citenamefont {Hsieh},
  \citenamefont {Lin}, \citenamefont {Liu}, \citenamefont {Duan}, \citenamefont
  {Bansil},\ and\ \citenamefont {Fu}}]{Fu:2012}%
  \BibitemOpen
  \bibfield  {author} {\bibinfo {author} {\bibfnamefont {T.}~\bibnamefont
  {Hsieh}}, \bibinfo {author} {\bibfnamefont {H.}~\bibnamefont {Lin}}, \bibinfo
  {author} {\bibfnamefont {J.}~\bibnamefont {Liu}}, \bibinfo {author}
  {\bibfnamefont {W.}~\bibnamefont {Duan}}, \bibinfo {author} {\bibfnamefont
  {A.}~\bibnamefont {Bansil}}, \ and\ \bibinfo {author} {\bibfnamefont
  {L.}~\bibnamefont {Fu}},\ }\href@noop {} {\bibfield  {journal} {\bibinfo
  {journal} {arXiv:1202.1003}\ } (\bibinfo {year} {2012})}\BibitemShut
  {NoStop}%
\bibitem [{\citenamefont {Alexandradinata}\ \emph {et~al.}()\citenamefont
  {Alexandradinata}, \citenamefont {Dai},\ and\ \citenamefont
  {Bernevig}}]{Alexandradinata:2012}%
  \BibitemOpen
  \bibfield  {author} {\bibinfo {author} {\bibfnamefont {A.}~\bibnamefont
  {Alexandradinata}}, \bibinfo {author} {\bibfnamefont {X.}~\bibnamefont
  {Dai}}, \ and\ \bibinfo {author} {\bibfnamefont {B.~A.}\ \bibnamefont
  {Bernevig}},\ }\href@noop {} {\bibinfo  {journal} {unpublished}\
  }\BibitemShut {NoStop}%
\bibitem [{\citenamefont {Resta}(1994)}]{Resta:1994}%
  \BibitemOpen
\bibfield  {journal} {  }\bibfield  {author} {\bibinfo {author} {\bibfnamefont
  {R.}~\bibnamefont {Resta}},\ }\href {\doibase 10.1103/RevModPhys.66.899}
  {\bibfield  {journal} {\bibinfo  {journal} {Rev. Mod. Phys.}\ }\textbf
  {\bibinfo {volume} {66}},\ \bibinfo {pages} {899} (\bibinfo {year}
  {1994})}\BibitemShut {NoStop}%
\bibitem [{\citenamefont {Essin}\ \emph {et~al.}(2010)\citenamefont {Essin},
  \citenamefont {Turner}, \citenamefont {Moore},\ and\ \citenamefont
  {Vanderbilt}}]{Essin2010}%
  \BibitemOpen
  \bibfield  {author} {\bibinfo {author} {\bibfnamefont {A.~M.}\ \bibnamefont
  {Essin}}, \bibinfo {author} {\bibfnamefont {A.~M.}\ \bibnamefont {Turner}},
  \bibinfo {author} {\bibfnamefont {J.~E.}\ \bibnamefont {Moore}}, \ and\
  \bibinfo {author} {\bibfnamefont {D.}~\bibnamefont {Vanderbilt}},\ }\href
  {\doibase 10.1103/PhysRevB.81.205104} {\bibfield  {journal} {\bibinfo
  {journal} {Phys. Rev. B}\ }\textbf {\bibinfo {volume} {81}},\ \bibinfo
  {pages} {205104} (\bibinfo {year} {2010})}\BibitemShut {NoStop}%
\bibitem [{\citenamefont {Malashevich}\ \emph {et~al.}(2010)\citenamefont
  {Malashevich}, \citenamefont {Souza}, \citenamefont {Coh},\ and\
  \citenamefont {Vanderbilt}}]{Malashevich2010}%
  \BibitemOpen
  \bibfield  {author} {\bibinfo {author} {\bibfnamefont {A.}~\bibnamefont
  {Malashevich}}, \bibinfo {author} {\bibfnamefont {I.}~\bibnamefont {Souza}},
  \bibinfo {author} {\bibfnamefont {S.}~\bibnamefont {Coh}}, \ and\ \bibinfo
  {author} {\bibfnamefont {D.}~\bibnamefont {Vanderbilt}},\ }\href
  {http://stacks.iop.org/1367-2630/12/i=5/a=053032} {\bibfield  {journal}
  {\bibinfo  {journal} {New Journal of Physics}\ }\textbf {\bibinfo {volume}
  {12}},\ \bibinfo {pages} {053032} (\bibinfo {year} {2010})}\BibitemShut
  {NoStop}%
\bibitem [{\citenamefont {Slager}\ \emph {et~al.}(2012)\citenamefont {Slager},
  \citenamefont {Mesaros}, \citenamefont {Juricis},\ and\ \citenamefont
  {Zaanen}}]{Slager2012}%
  \BibitemOpen
  \bibfield  {author} {\bibinfo {author} {\bibfnamefont {R.}~\bibnamefont
  {Slager}}, \bibinfo {author} {\bibfnamefont {A.}~\bibnamefont {Mesaros}},
  \bibinfo {author} {\bibfnamefont {V.}~\bibnamefont {Juricis}}, \ and\
  \bibinfo {author} {\bibfnamefont {J.}~\bibnamefont {Zaanen}},\ }\href@noop {}
  {\bibfield  {journal} {\bibinfo  {journal} {unpublished}\ } (\bibinfo {year}
  {2012})}\BibitemShut {NoStop}%
\end{thebibliography}
%

\end{document}